\documentclass[a4paper,11pt]{article}
\pdfoutput=1 

\usepackage{jheppub} 

\usepackage[T1]{fontenc} 
\usepackage{color}

\usepackage{amssymb}
\usepackage{slashed}
\usepackage[normalem]{ulem}

\newcommand{\alp}{a}
\newcommand{\ga}{g_{a\gamma}}
\newcommand{\photontheta}{\vartheta}

\title{Light in the beam dump\\ --\\ {\large Axion-Like Particle production from decay photons in proton beam-dumps}}

\author[a]{Babette D\"obrich}
\author[b]{Joerg Jaeckel}
\author[c]{and Tommaso Spadaro}

\affiliation[a]{CERN, Esplanade des Particules 1, 1211 Geneva 23 (Switzerland)}
\affiliation[b]{Institut f\"ur theoretische Physik, Universit\"at Heidelberg, \\Philosophenweg 16, 69120 Heidelberg (Germany)}
\affiliation[c]{Laboratori Nazionali di Frascati INFN, Via E. Fermi 40, 00044 Frascati (Italy)}

\emailAdd{babette.dobrich@cern.ch}
\emailAdd{jjaeckel@thphys.uni-heidelberg.de}
\emailAdd{tommaso.spadaro@cern.ch}

\abstract{
The exploration of long-lived particles in the MeV--GeV 
region is a formidable task but it may provide us a unique access to dark sectors. Fixed-target facilities with sufficiently energetic and intense proton beams are an ideal tool for this challenge.
In this work we show that the production rate of Axion-Like-Particles (ALPs)
coupled pre-dominantly to photons receives a significant contribution from daughter-photons
of secondary $\pi^0$ and $\eta$ mesons created in the proton shower. 
We carefully compare the PYTHIA simulated spectra of such secondaries to experimental literature,
compute the ALP flux
from the Primakoff conversion of these photons,
and finally revisit existing limits on ALPs and update the prospects for a set of existing and future searches.
Our results show that taking this production mechanism into account
significantly enhances the sensitivity compared to previous studies based on coherent ALP production 
in primary proton-nucleus interactions.
}

\begin{document} 
\maketitle
\flushbottom

\section{Introduction}
\label{sec:intro}

At present, the search for new particles
is driven by a number of pertinent observations that cannot be explained within the Standard Model (SM). Perhaps most prominent amongst them is Dark Matter (DM), but baryogenesis as well as the strong CP problem also ask for a solution.
At the same time, we are challenged by the (so-far) non-detection of new particles.
This could be explained by the new particles being extremely weakly coupled. If this is the case they can escape constraints from searches with colliders. But their small coupling also makes such particles long-lived.

In particular, very weakly coupled particles with masses in the MeV--GeV range provide exciting phenomenology
and interesting connections to Dark Matter
(cf., e.g.,~\cite{Boehm:2003hm,ArkaniHamed:2008qn,Freytsis:2010ne,Dienes:2013xya,Berlin:2015wwa,Alekhin:2015byh,Dolan:2017osp,Hochberg:2018rjs}) but also to baryogenesis~\cite{Akhmedov:1998qx,Asaka:2005pn,Shaposhnikov:2008pf} and perhaps even to the strong CP problem~\cite{Alves:2017avw,Berezhiani:2000gh,Agrawal:2017cmd}.
Consequently, Heavy Neutral Leptons, Dark Photons and pseudo-scalars
of such masses have received increasing attention over the past years (cf., e.g.,~\cite{Beacham:2019nyx} for a recent overview).
The latter -- pseudo-scalars -- also known as `Axion-like particles' (ALPs) --
are the subject of the present document.

ALPs have recently received a considerable amount
of interest in the context of Dark Matter model-building. 
They may act as a mediator
for the interactions between DM and SM particles and thereby allow
reproducing the correct Dark Matter relic abundance via thermal freeze-out. At the same time
this helps evading the strong constraints from direct and indirect
detection experiments~\cite{Freytsis:2010ne,Dienes:2013xya,Berlin:2015wwa,Dolan:2017osp}.
ALPs detectable with masses in the MeV-GeV range have also recently been discussed in the context of inflation~\cite{Takahashi:2019qmh}.
Other motivations for ALPs with masses above ${\mathcal{O}}(1)\,{\rm MeV}$
include their potential connection to
explaining  the observed value of the magnetic
moment of the muon~\cite{Marciano:2016yhf}. Also, via the so-called `relaxion' mechanism,
ALPs may play a crucial role in electroweak symmetry
breaking~\cite{Graham:2015cka} and in the solution of the hierarchy problem.
A concrete implementation~\cite{Flacke:2016szy} of such a relaxion  may yield
signatures that are observable with the methods described in this paper.
Moreover, it is worth reiterating that recently models have been proposed that allow the QCD axion to live at the MeV-GeV scale~\cite{Alves:2017avw,Berezhiani:2000gh,Agrawal:2017cmd}.
Beyond these phenomenological considerations, ALPs are also motivated by top-down extensions of the SM such as string theory~\cite{Svrcek:2006yi,Arvanitaki:2009fg,Acharya:2010zx,Cicoli:2012sz}. The crucial feature is that a weakly broken or anomalous shift symmetry allows their mass to be much smaller than the fundamental scale, making them accessible to experimental tests. The same shift symmetry then also ensures that the interaction is suppressed by the fundamental scale and therefore very weak.
In summary, new searches for ALPs are a well-motivated and timely task.
\bigskip

The main aim of the present paper is to study the production of ALPs in proton beam dumps from decay photons of secondary mesons.
We therefore focus on pseudoscalar ALPs  whose dominant
interaction is with photons. We use the Lagrangian
\begin{equation}
\mathcal{L}= \frac{1}{2} \partial^\mu a \, \partial_\mu a - \frac{1}{2}m^{2}_{\alp} \, \alp^2-\frac{1}{4} \, \ga \, \alp \, F^{\mu\nu}\tilde{F}_{\mu\nu} \; ,
\end{equation}  	
where $\ga$ is the photon-ALP coupling and $F^{\mu\nu}$ is the electromagnetic field strength.
\bigskip

To discover ALPs with masses on the order
of MeV-GeV, proton-fixed target facilities (or rather proton beam-dump experiments)
are well suited\footnote{Indeed there is a sizable number of such beams around the world, cf.~e.g.~\cite{Shiltsev:2013zma,Shiltsev:2014jpa,Shiltsev:2017mle,Alemany:2019vsk} whose suitability for ALP searches has recently been discussed in~\cite{Harland-Lang:2019zur}.}.
The strength of such an experimental setup is that it can provide sufficient energy to produce MeV-GeV scale particles, while ensuring that all of the protons in the beam ultimately interact. Moreover, decay volumes spanning tens of meters allow ALPs of various lifetimes to be detected.
Overall this combination provides high sensitivity but also excellent complementarity to experiments at low-energy colliders, 
such as Belle-II~\cite{Dolan:2017osp}, which can explore the region below a ${\rm few}\,{\rm GeV}$ but relatively strong coupling, as well as to experiments at the LHC, which are sensitive mostly to masses 
above a ${\rm few}\,{\rm GeV}$~\cite{Jaeckel:2012yz,Mimasu:2014nea,Jaeckel:2015jla,Mariotti:2017vtv}\footnote{Missing energy signatures could give access to ALPs with lower mass also at LHC. For a recent study see, e.g.~\cite{Ebadi:2019gij}.}.

For ALPs coupled predominantly to photons,
there are (at least) two important production mechanisms to be considered.
In both cases, one of the photons is provided by 
a nucleus at rest, constituting the target/dump material. The second photon
can be provided
\begin{enumerate}
 \item by the charged proton itself. This is sometimes referred to as `photon-from-proton' (PFP) mode,
 where the photon distribution around the proton is often computed in the Weizsaecker-Williams approximation\footnote{More precisely usually variants of the equivalent photon approximation~\cite{Budnev:1974de} are used~\cite{Dobrich:2015jyk,Dolan:2017osp}. In particular in the region of low ALP masses this approximation is problematic and receives sizable corrections that lower the cross section~\cite{Harland-Lang:2019zur}.
 This further increases the relative importance of the production mechanism that we will discuss in the present paper.}
 \item through a decay, notably from $\pi^0 \rightarrow \gamma \gamma$, but also
 from other neutral mesons.
\end{enumerate}
As already noted, the second process is the main focus of our paper.

\begin{figure}[]
\begin{center}
 \includegraphics[width=0.6\textwidth]{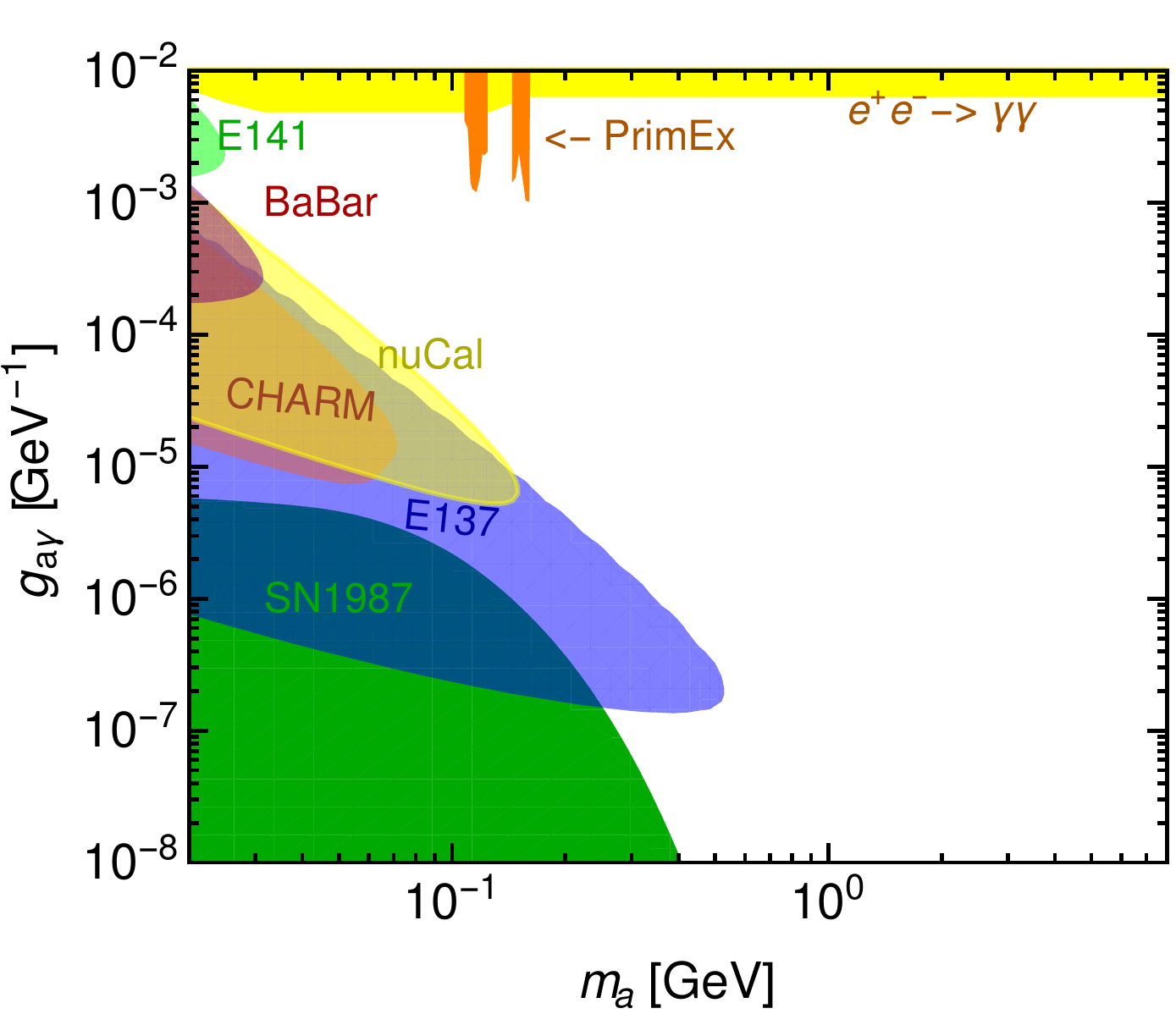}
 \caption{\label{fig:ALPs_status}
 Status of exclusions for ALPs coupled to photons in the MeV-GeV range.
 The limits are taken from \cite{Dolan:2017osp} updated with the PrimEx recast~\cite{Aloni:2019ruo}.
}
 \end{center}
\end{figure}

Although the importance
of the inclusion of axion
production from secondary pions, was, for example,
already pointed out in~\cite{Tsai:1986tx}, it is still somewhat under-appreciated.
Indeed all the proton beam dump constraints shown in the overview Fig.~\ref{fig:ALPs_status} are calculated
using only the PFP production mode~\cite{Dobrich:2015jyk,Dolan:2017osp}.
To our knowledge, only\footnote{
A recent study performed for the FASER experiment~\cite{Feng:2018pew} at LHC
includes the estimate of ALP production directly from $\pi^0$
decays  (which they find subdominant) as well as Primakoff-produced ALPs from $\pi^0$ decays, 
albeit in a different kinematic regime. } a study put forward in~\cite{Berlin:2018pwi} which determined prospects
for SeaQuest (after its proposed ECAL upgrade) takes into account an estimate
of the ALP yield stemming from the Primakoff-conversion of
photons from $\pi^0$ decays in the dump\footnote{While also considering Primakoff production from  
real photons the PrimEx and GlueEx experiments considered in~\cite{Aloni:2019ruo} are effectively a photon fixed-target experiment where the photons are produced via Bremsstrahlung off an electron beam and then shot onto the target.}. However, some simplifying assumptions are made and no full Monte Carlo is set-up.

With our work, we want to close this gap
in the literature and give improved estimates for the ALP production from meson-decay photons. The sensitivity improvement with respect to the case when only the PFP production is included will be discussed.

Our study is particularly timely
since some of the experiments that can impact the 
parameter space have started taking data, notably NA62 \cite{NA62:2017rwk}, or are close to data-taking.

In practice calculating the photon flux inside the beam dump is far from trivial due to the non-perturbative nature of meson production. We therefore first carefully
compare the yields for $\pi^0$ and other meson and their related angular distributions from PYTHIA~\cite{Sjostrand:2014zea} simulations
to data from past experiments and then use it to determine the photon flux inside the dump.

While here we are interested mostly in ALP production,
the $\pi^0$ and other meson spectra in the dump are not only of relevance to the
production of ALPs but can also be the source of Dark Photons and 
other exotic particles. Thus our work in comparing the yield
from PYTHIA simulations to data is of more general interest.

Our paper is structured as follows. A first direct comparison of the PYTHIA simulation output with the experimental 
data is performed in Section~\ref{sec:yield}: proton-proton 
and proton-beryllium interactions are separately analyzed.
In Section~\ref{sec:theory} we review and discuss the computation of the ALP yield 
through production from the meson-decay photons.
Finally, in Section~\ref{sec:sensi} we re-evaluate existing experimental constraints and make estimates for future sensitivities taking the additional production mechanism into account. We discuss the conclusions in Section~\ref{conclusions}.

\bigskip
{\bf Note added:} 
Appendix~\ref{app:erratum} was added after publication of this work in JHEP (cf.~\cite{Dobrich:2019dxc}) and corrects\footnote{We are very grateful to F.~Kahlhoefer and his group for noting the discrepancy that led to the discovery of the error in our code.} some of the figures in the main text.

\section{Neutral meson yields in proton beam dumps} \label{sec:yield}
The simulation of the production rates of secondary mesons is a challenging task, since 
the formation of mesons is complicated due to non-perturbative physics. In this section we therefore validate our PYTHIA simulations with experimental data. While no measurements of inclusive neutral meson production are available at exactly the desired energies and target materials employed in the experiments considered in Section~\ref{sec:sensi}, we nevertheless have data covering the energy range from 60~GeV to 450~GeV and different target materials, in particular hydrogen and beryllium. Furthermore we can compare to the production of different types of mesons. Putting this together allows us to have at least some confidence in the employed meson spectra. The total cross section for inclusive $\pi^0$ production agrees with the data within an uncertainty of 20\% in the entire beam momentum range of interest.  We also indicate kinematic regions where the results are more uncertain. Including and neglecting the contributions from these regions we provide an estimate of the uncertainty of the limits and sensitivities in Section~\ref{sec:sensi}.
 
\subsection{400 GeV proton beam on a hydrogen target}
Measurements of secondary meson production from a $400\,{\rm GeV}$ proton beam dumped onto a hydrogen target have been performed at the 
beginning of the 1990's by the NA27 experiment operating at the LExan Liquid hydrogen Bubble Chamber (LEBC) with the European Hybrid Spectrometer (EHS). 
Results for $\pi^0$ and $\eta$ production from the LEBC-EHS~\cite{AguilarBenitez:1991yy} allow a direct comparison of the proton-proton interaction expectation from the 
PYTHIA simulation program~\cite{Sjostrand:2014zea} with the experimental results. 
The simulation includes elastic, inelastic non diffractive, and single-, double-diffractive processes. Parton densities for protons are defined using the CTEQ 5L set~\cite{Lai:1999wy}, a widely-used leading-order QCD parametrization with $\alpha_s(M_Z) = 0.127$. 

The measurements from LEBC-EHS report that $\pi^0$ ($\eta$) are produced with an average multiplicity of $3.87\pm0.12$ ($0.30\pm0.02$) per incident proton.
A total of $51.2\pm3.1~\%$ of the produced $\pi^0$s stem from the decay of secondary particles (mostly mesons).
These figures can be compared with the output of the PYTHIA simulation. From it, the total $\pi^0$ ($\eta$) production multiplicity is $4.248\pm0.007$ ($0.489\pm0.002$). 
A total of $48.1\pm0.7~\%$ of the produced $\pi^0$s stem from the decay of $\rho^\pm$, $\rho^0$, $\omega$, or $\eta$, in good agreement with the experimental data. 
For the total cross section, the SoftQCD set of PYTHIA version 8.2 yields 39.9~mb summing up single- and double-diffractive, non-diffractive, and elastic processes, in good agreement with the experimental data which slightly exceeds 40~mb~\cite{ppPDG}.

\begin{figure}
\begin{center}
 \includegraphics[width=0.49\textwidth]{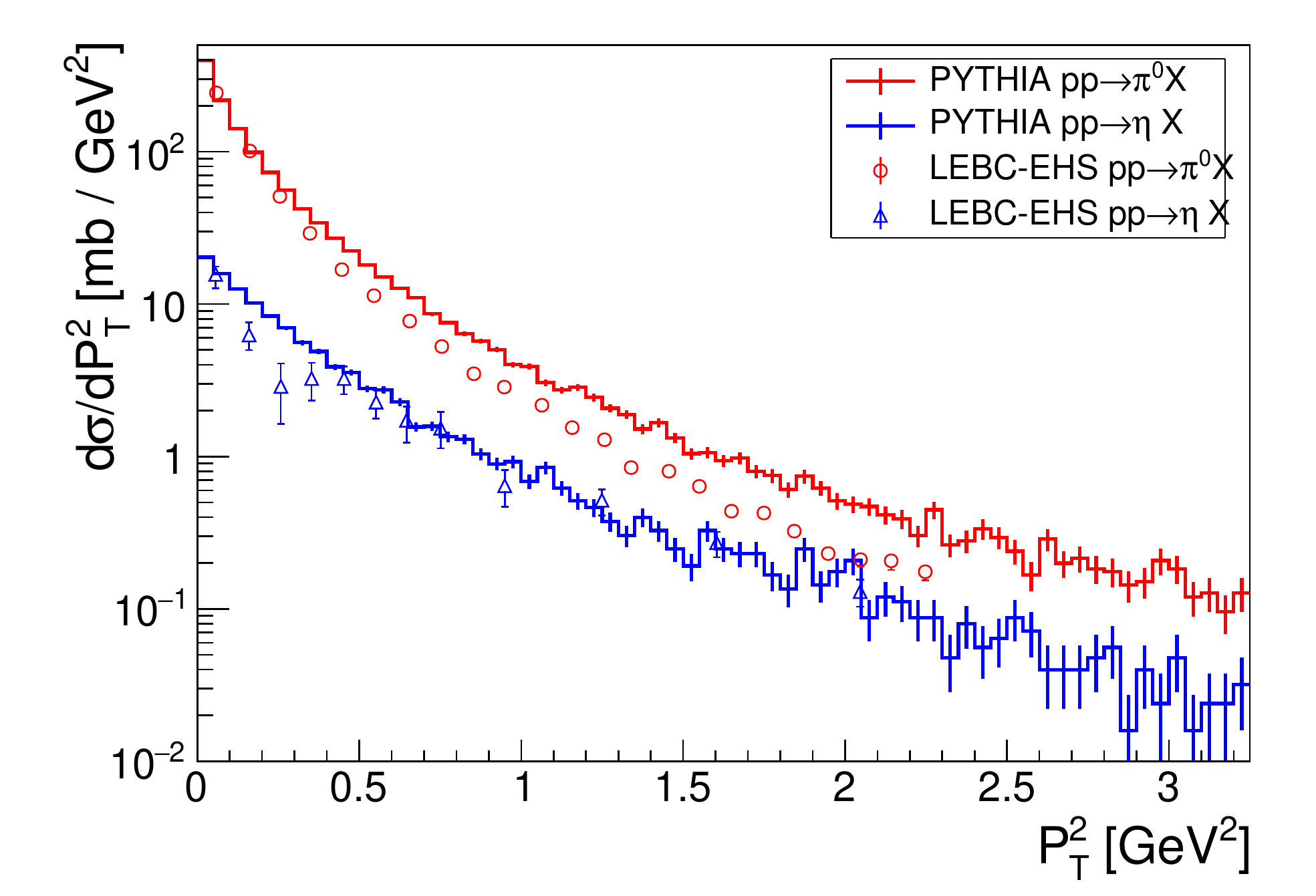}
 \includegraphics[width=0.49\textwidth]{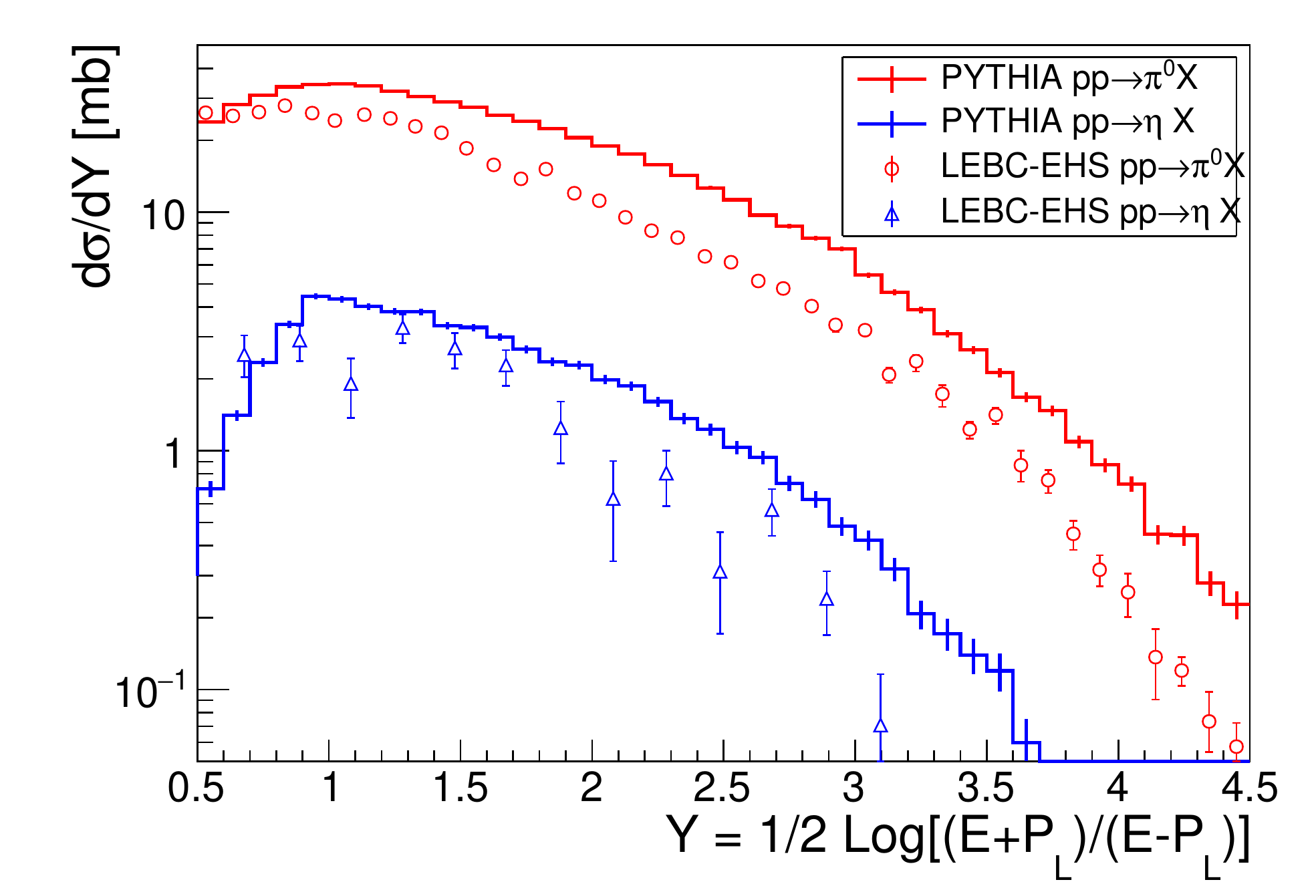}
 \caption{Measurements from scattering of 400 GeV protons onto a hydrogen target for $\pi^0$ (red) or $\eta$ mesons (blue), compared with the expectation from 
 PYTHIA 8.2: differential cross sections as a function of the squared transverse momentum (left panel) and rapidity (right panel).
Symbols refer to results from the NA27 experiment at the LEBC-EHC. The histograms are the output of a PYTHIA MC simulation.
\label{fig:LEBC_vs_PYTHIA}
 }
 \end{center}
\end{figure}

\begin{figure}
\begin{center}
 \includegraphics[width=0.60\textwidth]{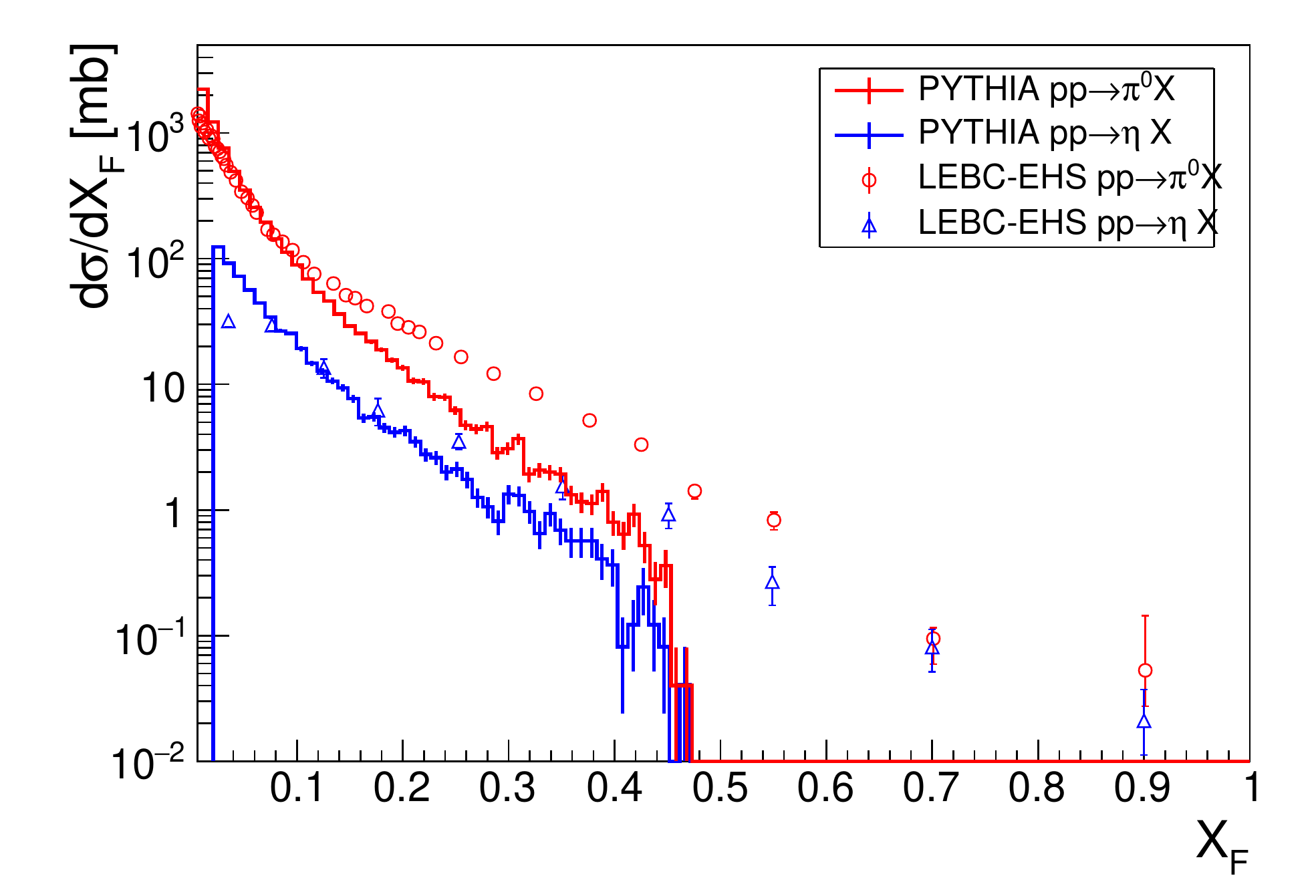}
 \caption{Measurements from scattering of 400 GeV protons onto a Hydrogen target for $\pi^0$ (red) or $\eta$ mesons (blue), compared with the expectation from 
 PYTHIA 8.2: differential cross section as a function of the Feynman $X_F$ variable.
Symbols refer to results from the NA27 experiment at the LEBC-EHC. The histograms are the output of a PYTHIA MC simulation.
\label{fig:LEBC_vs_PYTHIA_xf}
 }
 \end{center}
\end{figure}

In Figs.~\ref{fig:LEBC_vs_PYTHIA} and~\ref{fig:LEBC_vs_PYTHIA_xf}, the distributions measured at LEBC-EHS for the squared transverse momentum ($P_T^2$), rapidity ($Y$), and Feynman $X_F$ variable have been compared to the PYTHIA output, after applying to the Monte Carlo the experimental selection criteria: for $\pi^0$ ($\eta$), the condition is $X_F > 0.006\mbox{ }(0.021)$, where the Feynman variable is computed as $X_F = P_Z/P_Z(\mathrm{max})\sim 2P^\ast_Z/\sqrt{s}$, $Z$ represents the beam axis direction, and $P^\ast$ is evaluated in the center of mass frame. 
To obtain the differential cross sections, the MC output is scaled according to the total proton-proton cross section, 39.14~mb for a 
$400\,{\rm GeV}$ proton beam.
Four regions of $X_F$ are defined: a central region, for $X_F < 0.025$, where the MC overestimates the data by a factor less than two; an intermediate region, for $0.025 < X_F < 0.1$, where data and MC agree; a fragmentation region, for $0.1 < X_F < 0.3$ where MC underestimates the data by a factor less than three, and a forward contribution, for $X_F>0.3$, associated to the inelastic diffraction mechanism, where the MC largely underestimates the data. Inserting in the simulation the double-pomeron exchange with the Minimum-Bias Rockefeller model~\cite{Ciesielski:2012mc} is seen to slightly improve the data-MC comparison for the central region, while not affecting the other regions. Including the double-pomeron exchange increases the total proton-proton cross section by 0.47~mb for a proton beam energy of 400~GeV. 

In the following, the $X_F$ domain has been divided into the eight bins defined in Tab.~\ref{tab:xfBins}. In general, the probability that photons from $\pi^0$ decays produce ALPs in the experimental acceptance increases with $X_F$. As discussed in Section~\ref{sec:theory}, we use our knowledge of the quality of our simulated meson spectra by making the following estimate for the uncertainty. As a baseline we take into account the full simulation results including all $X_F$ bins (0-7). This can then be compared to a conservative estimate that only includes bin 5, where the agreement between simulation and data is very good, and bin 6 where the simulation underestimates the data by a moderate factor of up to 3.

\begin{table}
\begin{center}
\begin{tabular}{ |c|| c|c|  c|c|   c|c|   c|c|  } 
 \hline
 $X_F$ bin & 0 & 1 & 2 & 3 & 4 & 5 & 6 & 7 \\ \hline
  $X_F$ min & -0.6 & -0.3 & -0.1 & -0.025 & 0 & 0.025 & 0.1 & 0.3 \\ \hline
  $X_F$ max & -0.3 & -0.1 & -0.025 & 0 & 0.025 & 0.1 & 0.3 & 0.6 \\ \hline
\end{tabular}
 \caption{\label{tab:xfBins} Definition of bins in the Feynman $X_F$ variable. For details, see text.}
 \end{center}
 \end{table}


\subsection{400 GeV proton beam on a beryllium target}
We can now take the next step and allow for different target materials. First we note our expectation that 400 GeV 
protons interacting with a fixed proton target do not differ significantly from those of a neutron target, 
as far as the $\pi^0$ or $\eta$ production is concerned. This is confirmed by the PYTHIA Monte Carlo both for the scattering distributions and cross sections. Following from this we obtain the cross sections for larger target nuclei by an appropriate scaling with the geometric cross section $\sim A^{2/3}$, which in this subsection we compare to the data.

Measurements of meson yields from a 400~GeV proton beam on beryllium targets have been performed in 
\cite{Atherton:1980vj}, and data was taken for four values of the secondary particle momenta (60, 120 and 300 GeV) and two values of transverse-momentum (0 and 500 MeV)
at different target lengths. To complement these measurements at a lower momentum range of secondary 
particles and in view of evaluation of neutrino fluxes
for NOMAD and CHORUS, the NA56/SPY experiment~\cite{Ambrosini:1999id} published yields in the
range of secondary momentum from 7 to 135 GeV with a proton beam of 450~GeV. In order to make these experimental data
useful  for further applications, a very useful parametrization was developed in~\cite{Bonesini:2001iz},
and sometimes is referred to as `BMPT'\footnote{In the context of the production of exotic 
particles, BMPT was, e.g., also employed in~\cite{deNiverville:2016rqh}
to predict yields of sub-GeV Dark Matter production in beam dump experiments.}.
As this parametrization was developed by the extrapolation of 400~GeV and 450~GeV data,
it is suited to be employed for the use-case of NA62 and SHiP~\cite{Alekhin:2015byh,Anelli:2015pba}, while care has to be taken, when 
extrapolating to the NuCal~\cite{Blumlein:1990ay}, and SeaQuest beam energies of 70~GeV and 120~GeV, respectively.

If the contribution of heavier meson and resonance decays is removed, the $\pi^0$ inclusive cross section is 
expected to be approximately equal to the average of the $\pi^+$ and $\pi^-$ inclusive cross sections. To validate our estimates for $\pi^0$ based on PYTHIA, we directly compare the MC output to the inclusive invariant
cross section $E \ {\rm d}^3 \sigma/ {\rm d}p^3$ obtained in~\cite{Ambrosini:1999id}
for $\pi^+$ and $\pi^-$. For completeness, we also consider the emission of $K^+$ and $K^-$, protons and anti-protons.

\begin{figure}[]
\begin{center}
 \includegraphics[width=0.49\textwidth]{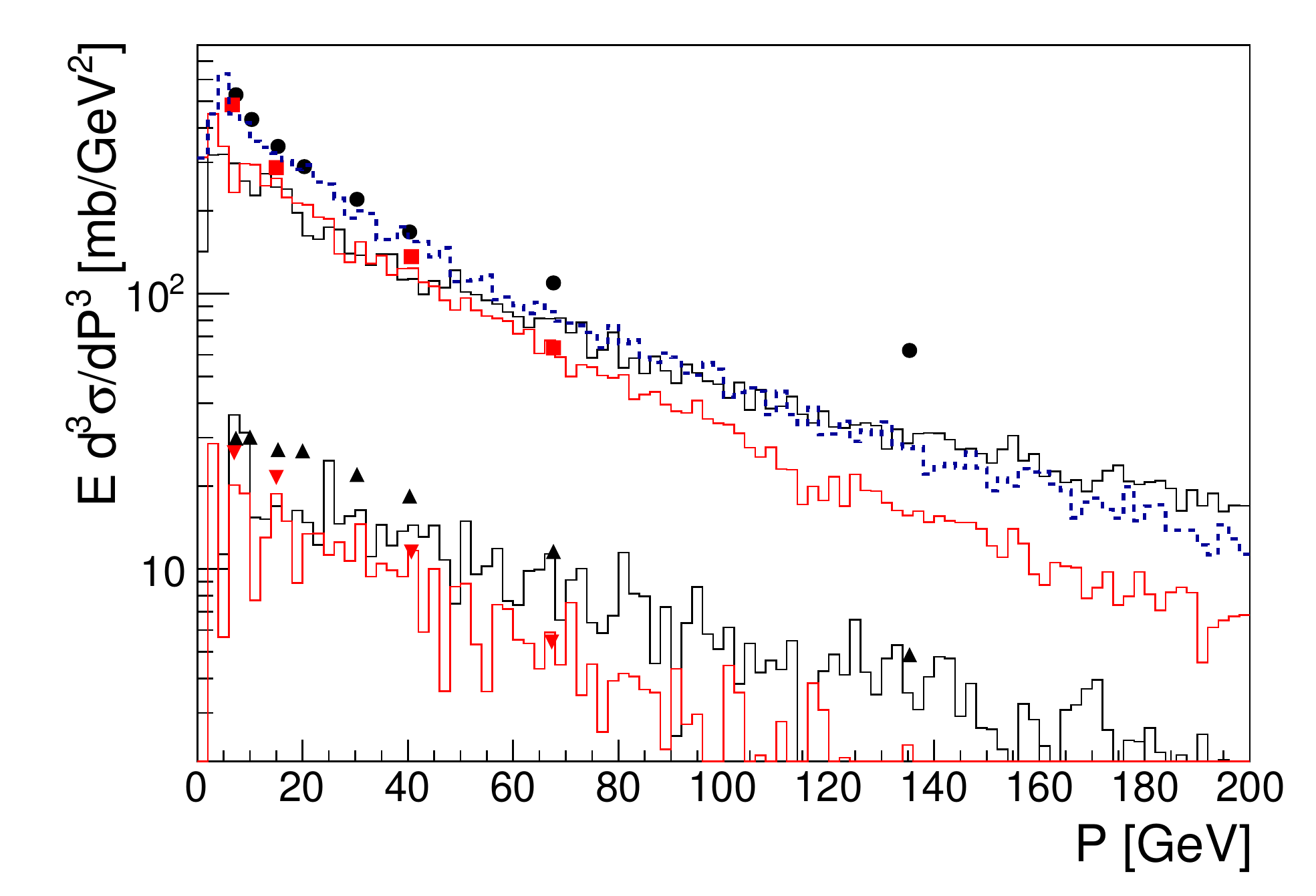}
 \includegraphics[width=0.49\textwidth]{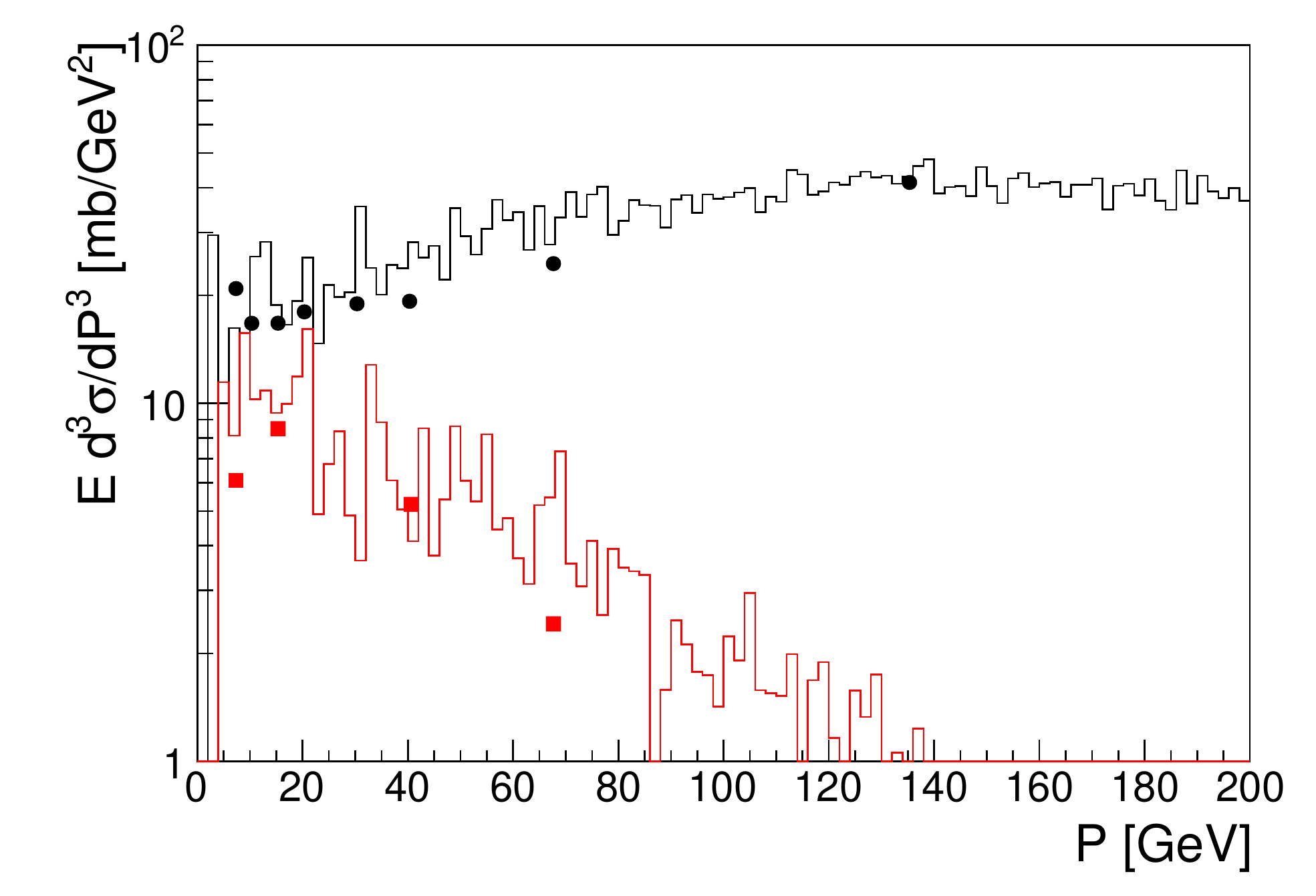}
  \caption{ Left-hand-side plot: Inclusive invariant cross-sections measured by NA56~ compared to the output of the PYTHIA simulation (code version 8.2)
 with the same acceptance cuts applied by the experiment. Symbols represent NA56 data~\cite{Ambrosini:1999id} (black dots for $\pi^+$, red squares for $\pi^-$), 
 black and red histograms represent the PYTHIA output for a 450~GeV proton beam. The dashed blue histogram indicates the expected inclusive invariant cross section for $\pi^0$ emission.
 The lower black and red histograms from PYTHIA refer to $K^+$ (black) and $K^-$ (red) and can be compared to the NA56 results for $K^+$ (black up-pointing triangles) 
 and $K^-$ (red down-pointing triangles). 
 Right-hand-side plot: Inclusive invariant cross sections for protons (antiprotons) from NA56 represented by black circles (red squares), to be compared to black (red)
 histograms from PYTHIA.
\label{fig:NA56_comparison}
 }
 \end{center}
 \end{figure}

Figure~\ref{fig:NA56_comparison} shows the results of our comparison, which is quite
good. 
To arrive at this comparison we have
\begin{enumerate}
\item accounted for the angular acceptance of NA56, corresponding to emission angles below approximately 0.7 mrad;
\item accounted for the NA56 target-efficiency factor: $\big( 1-\exp ( L/\lambda_{\rm Be} ) \big) $ (where $L$ is the target length and $\lambda_{\rm BE}$ is the appropriate proton interaction length)
\item scaled the MC results for the proton-beryllium total cross section, a factor of $A^{2/3}$ higher than that for proton-proton scattering.
\end{enumerate}
Above, we have assumed a target length $L=100$~mm and a proton interaction length $\lambda_{\rm Be}=423$~mm.
In Figure~\ref{fig:NA56_comparison}, left-hand side, we observe a very satisfying agreement
for all available data from NA56 except a slight underestimation of the yield 
of $\pi^+$ at large momenta.  As NA56 data might include tertiary production up to a certain extent, an 
underestimation with respect to the simulation output might be expected.
On the right-hand-side of Figure~\ref{fig:NA56_comparison}, for completeness
we also show the proton and anti-proton inclusive invariant cross sections measured at NA56 compared to the expected output from 
PYTHIA.\footnote{The agreement observed on the secondary production would allow a reliable estimate of the $\pi^0$ tertiary production in a dump, which can be expected to be not negligible with respect to the primary production. This evaluation depends on the detailed structure of the dump and is beyond the scope of the present paper.}

\subsection{Proton beam energies below 400 GeV}
\label{Sec:lowE}
\begin{figure}[!t]
\begin{center}
 \includegraphics[width=0.49\textwidth]{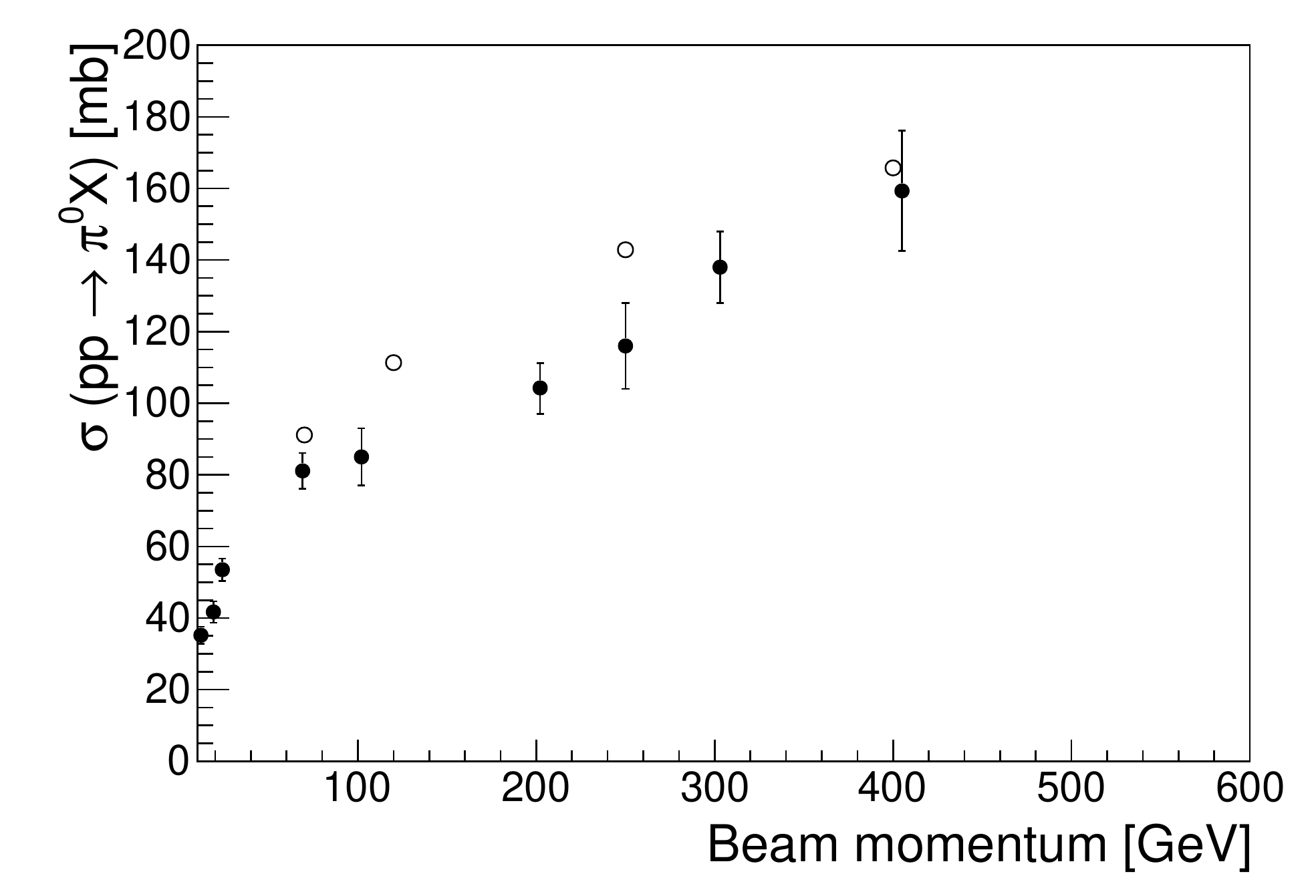}
 \includegraphics[width=0.49\textwidth]{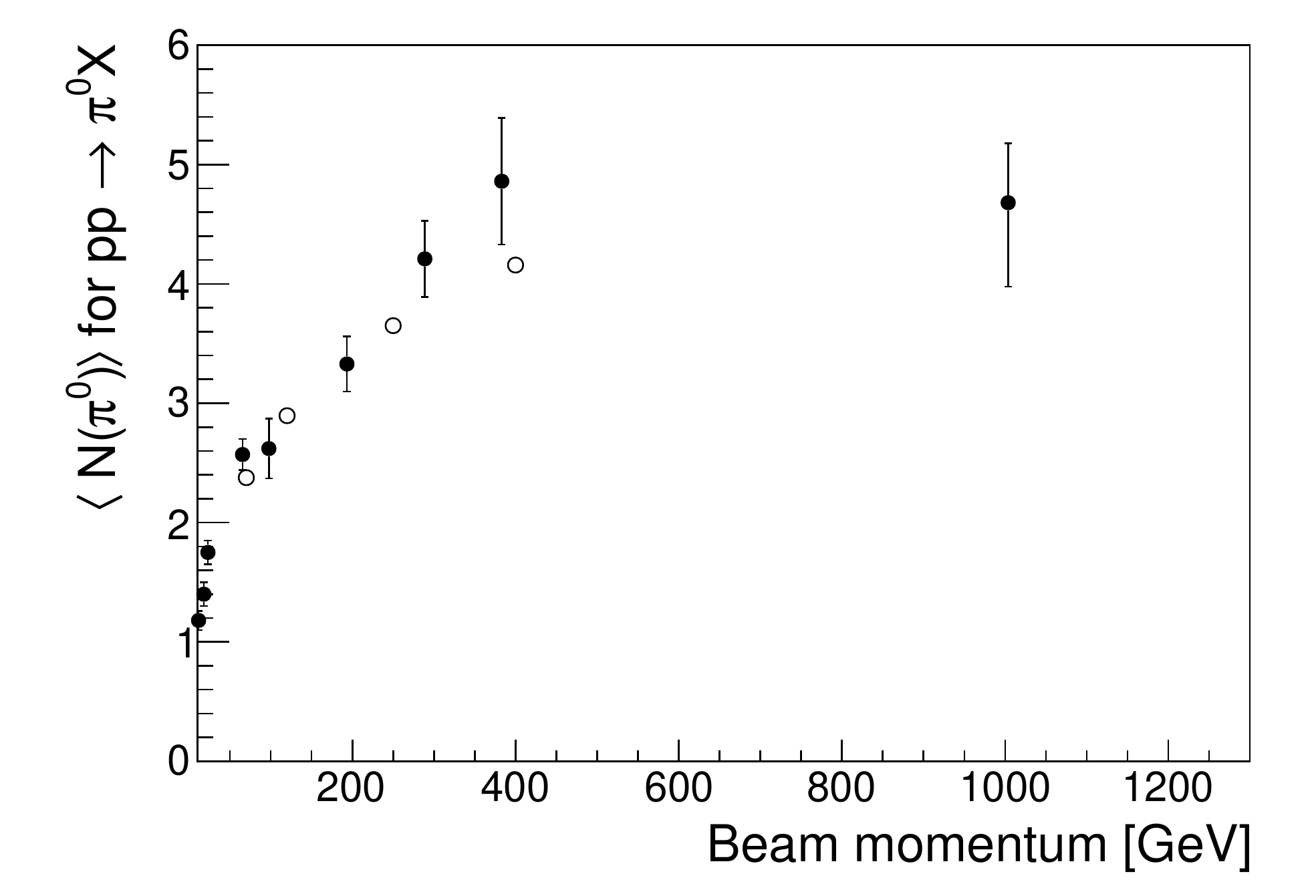}
  \caption{Left-hand-side plot: the total cross-section measured for inclusive $\pi^0$ production (filled dots) is compared to the output of the PYTHIA simulation, code version 8.2 (open dots). 
 Right-hand-side plot: $\pi^0$-meson average yield from data (filled dots)~\cite{sigmatotLiterature} are 
compared to the expectation from PYTHIA (open dots). 
\label{fig:xsmult}
 }
 \end{center}
 \end{figure}

To compare the inclusive production of light mesons with available literature, PYTHIA simulations of proton-proton interaction have been produced for proton beam energies of 70, 120, and 250~GeV: the first two values correspond to the beam energy of 
the NuCal and SeaQuest experiments, while the third value corresponds to the NA22 experiment~\cite{Adamus:1986ta}, providing the most complete experimental data available below 400~GeV.  The total cross section of $\pi^0$ production and the average number of emitted $\pi^0$ mesons are shown in Fig.~\ref{fig:xsmult} as a function of the beam energy: a general agreement (within 20\% relative uncertainty) is observed between data~\cite{sigmatotLiterature} and MC.  

\begin{figure}[!t]
\begin{center}
 \includegraphics[width=0.49\textwidth]{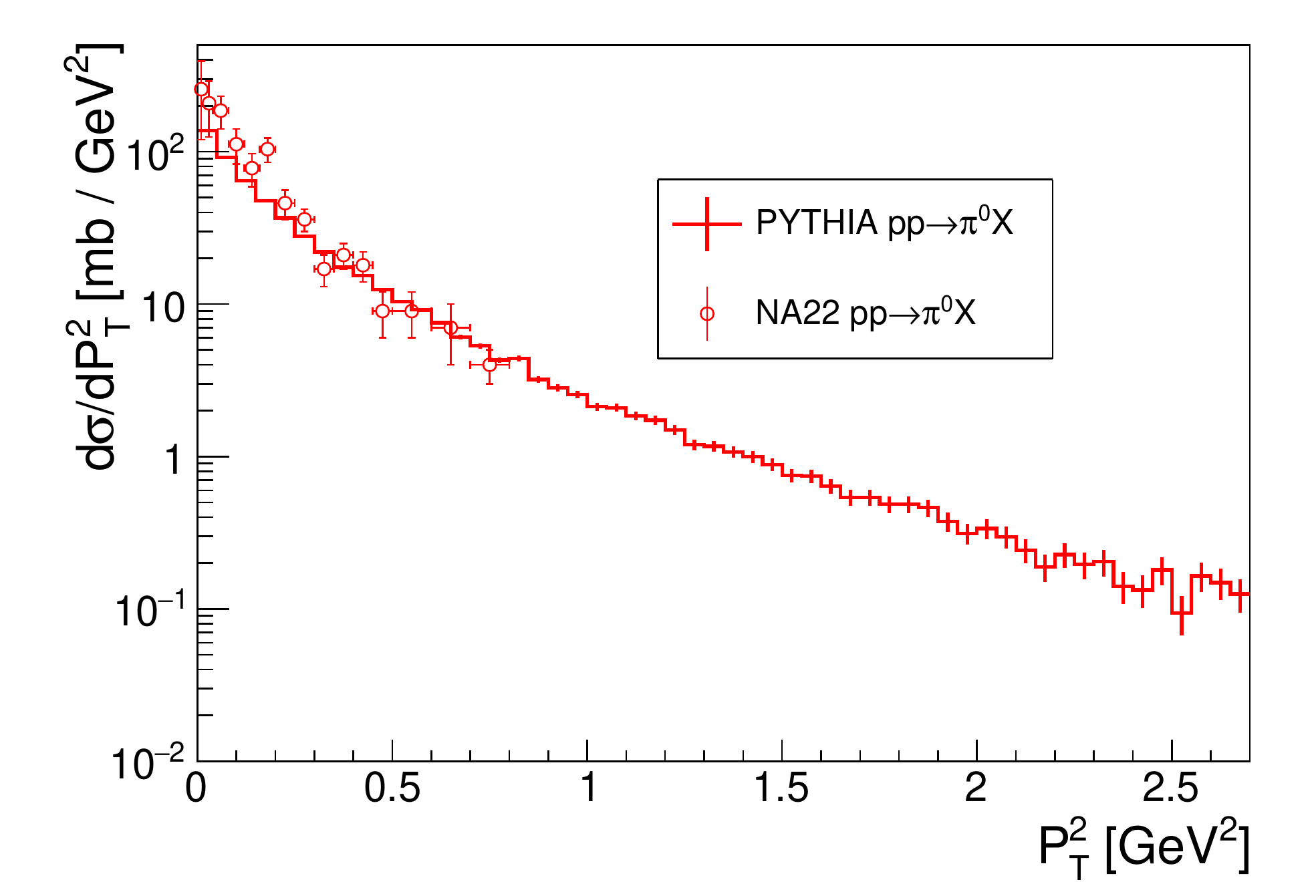}
 \includegraphics[width=0.49\textwidth]{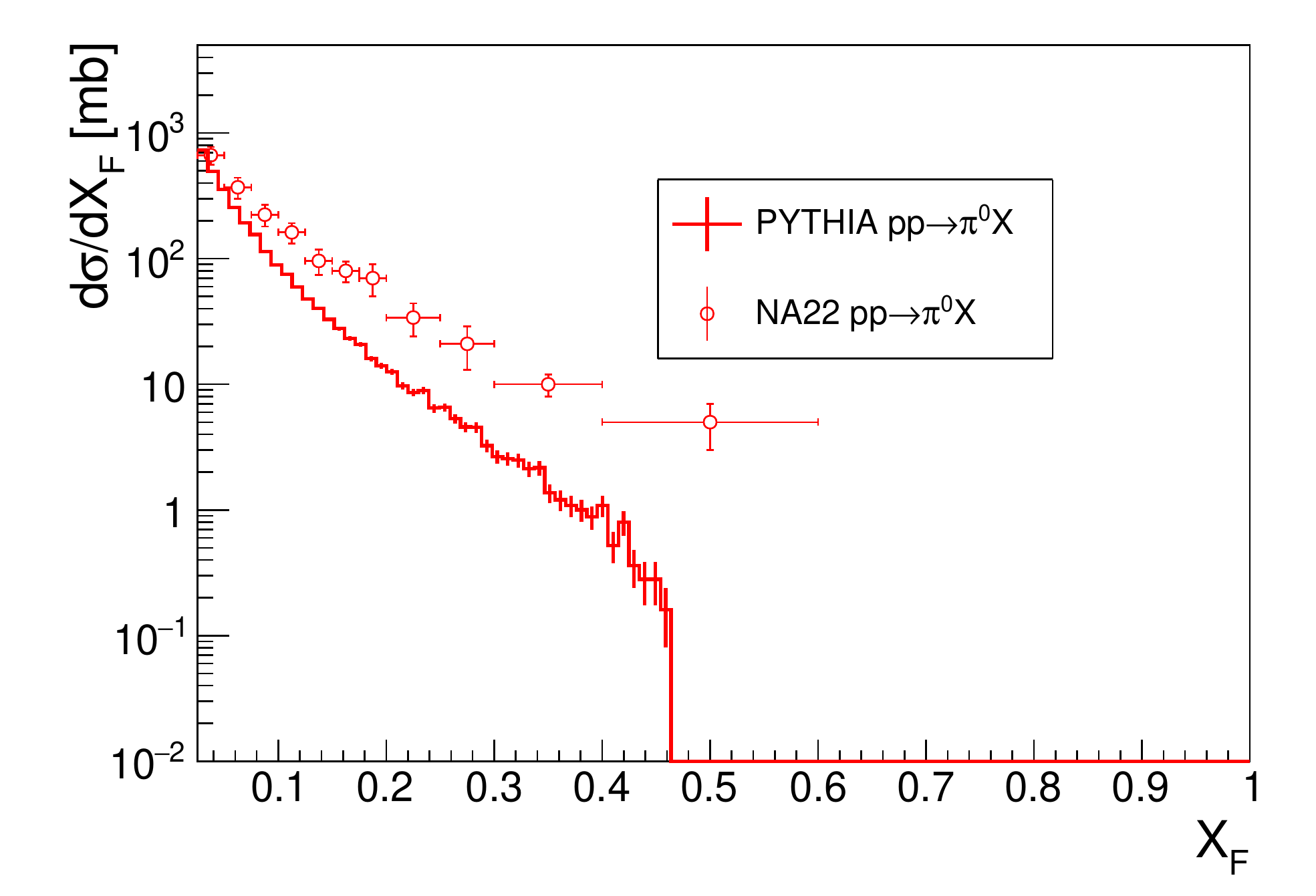}
 \caption{Measurements from scattering of 250 GeV protons onto a hydrogen target for $\pi^0$ mesons, compared with the expectation from 
 PYTHIA 8.2: Differential cross section as a function of the squared transverse momentum (left panel) and Feynman $X_F$ variable (right panel).
 Filled symbols refer to results from the NA22 experiment at the European Hybrid Spectrometer; open symbols are the output of a PYTHIA MC simulation.
\label{fig:NA22_vs_PYTHIA_dsigma}
 }
 \end{center}
\end{figure}

\begin{figure}[!t]
\begin{center}
 \includegraphics[width=0.6\textwidth]{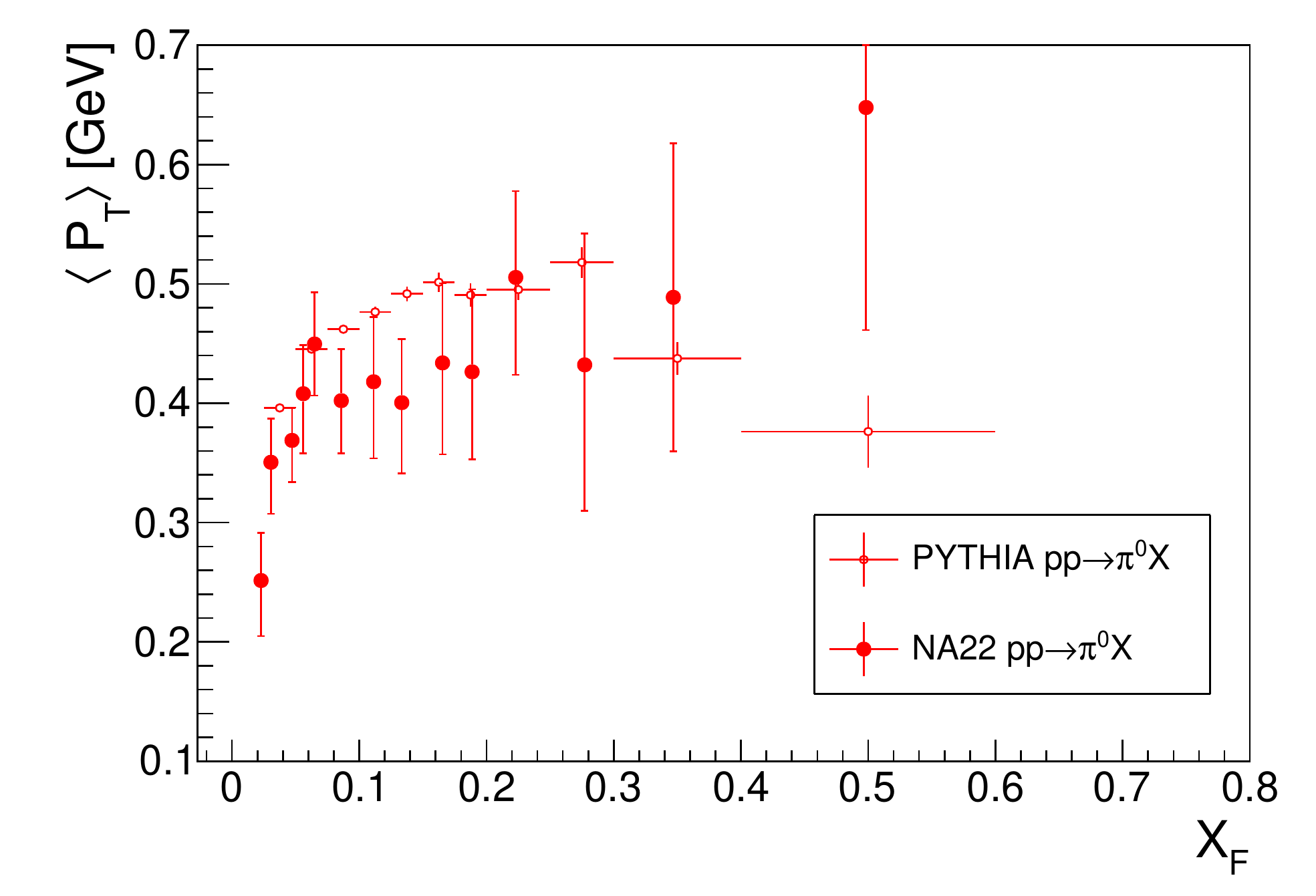}
\caption{Measurements from scattering of 250 GeV protons onto a hydrogen target for $\pi^0$ mesons, compared with the expectation from 
 PYTHIA 8.2: Correlation between the average transverse momentum and $X_F$ (right panel).
 Filled symbols refer to results from the NA22 experiment at the European Hybrid Spectrometer; open symbols are the output of a PYTHIA MC simulation.
\label{fig:NA22_vs_PYTHIA_corr}
} 
 \end{center}
\end{figure}

Data and MC differential cross sections have been compared at 250~GeV, as a function of the squared transverse momentum $P_{T}^{2}$ and of the Feynman $X_F$ variable (Fig.~\ref{fig:NA22_vs_PYTHIA_dsigma}). Data and MC transverse momentum differential cross sections are seen to agree (even if the data range is limited), while MC underestimates the data for $X_F>0.05$. Again, this will lead to conservative estimates of the expected ALP yields in the following. The correlation between average transverse momentum and $X_F$ observed in data is quite well reproduced by the MC, as shown in Fig.~\ref{fig:NA22_vs_PYTHIA_corr}.

\section{Production of ALPs from meson decay photons\label{sec:theory} }

Having validated our PYTHIA spectra for neutral secondaries
produced in proton interactions against data in the previous section~\ref{sec:yield}, we are now
ready to compute their impact on ALP production.
We will also compare it with the contribution of ALPs from the photon-from-proton-mode,
which has been evaluated in Ref.~\cite{Dobrich:2015jyk}.
As both processes have different initial and final states 
there is no interference and both contributions can be added together. In the following we will therefore concentrate on the new contribution from the mesons.

Taking the meson distributions as an input we set up a Monte-Carlo simulation that proceeds along the following steps:
\begin{enumerate}
 \item Simulate the decay of the neutral mesons produced in the dump into photons.
 \item Compute the cross-section of ALP production from these photons in the target nucleus.
 \item Mimic appropriate experimental acceptances and cuts.
 \item Evaluate a sensitivity prospect at fixed number of incident protons for the
 situation of zero background.
\end{enumerate}

The decay length of the neutral pion is given by
\begin{equation}
\ell^{\rm decay}_{\pi^{0}}=0.02\,{\rm mm}\left(\frac{E_{\pi^{0}}}{100\,{\rm GeV}}\right).
\end{equation}
The decay length of the other neutral mesons is even smaller.
Therefore, effectively the meson decay into photons is instantaneous, i.e. all mesons decay inside the target. 

\begin{figure}[]
\begin{center}
 \includegraphics[width=0.49\textwidth]{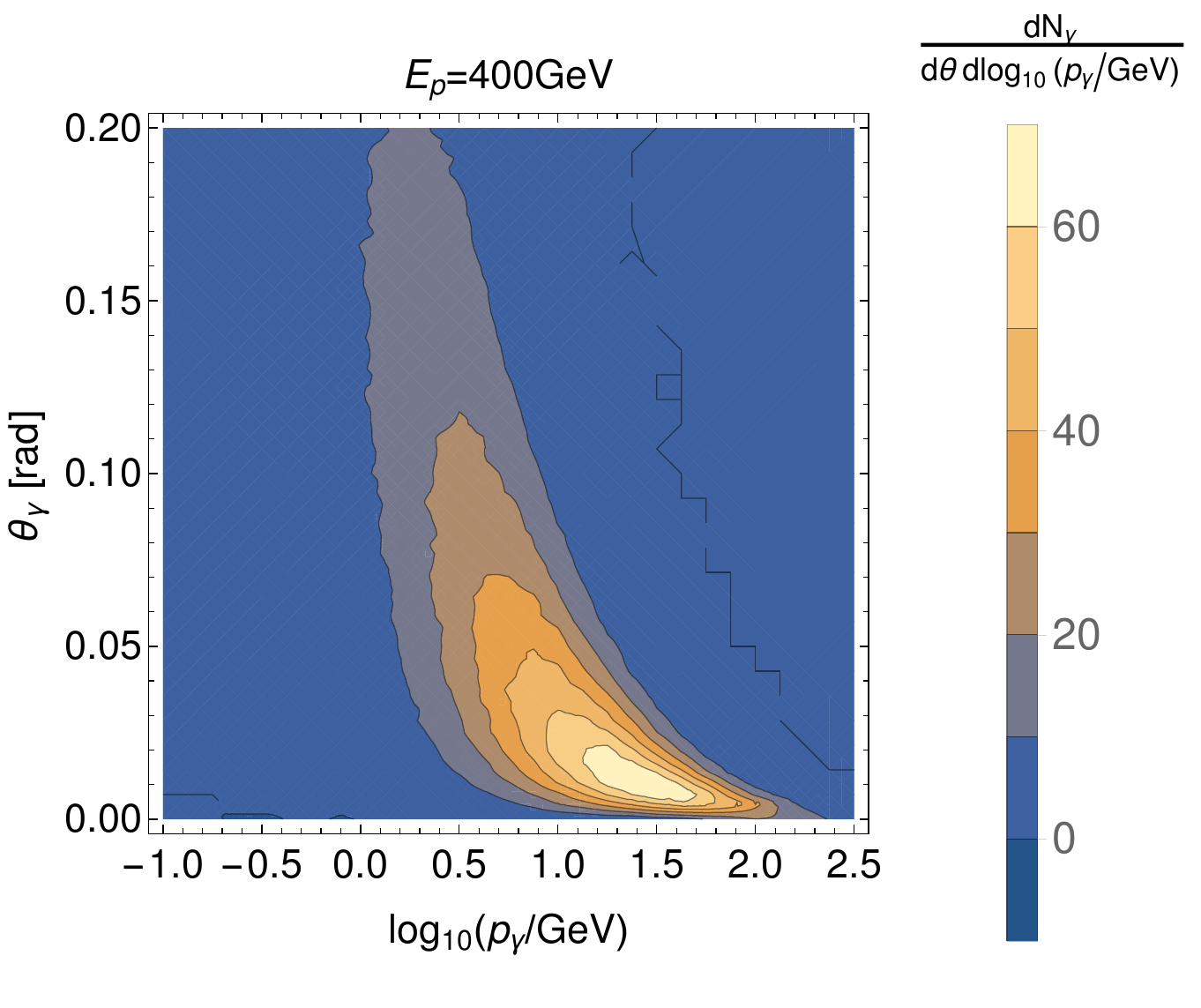}
  \includegraphics[width=0.49\textwidth]{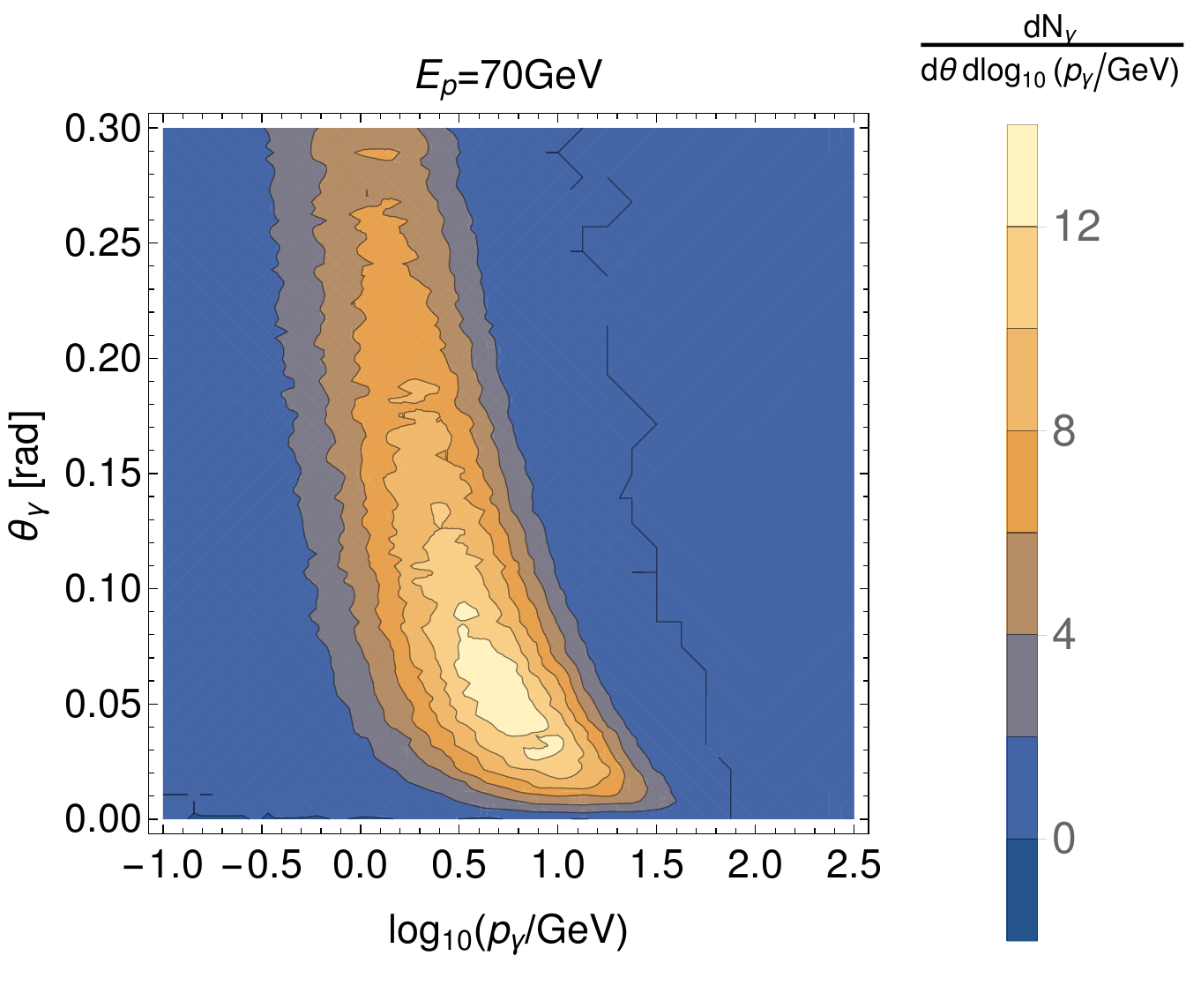}
 
 \caption{Distribution $\theta$ vs $\log_{10}(p_\gamma)$ of photons from $\pi^0$ decay
 for all $X_F$ bins normalized to the number of simulated proton events $N_{\it sim}=10^5$. 
 Left: incident proton beam at $400\,{\rm GeV}$.
 Right: incident proton beam at $70\,{\rm GeV}$. The integrated number of photons per
 incoming proton is $\sim 9$ for $400\,{\rm GeV}$ and $\sim 5$ for $70\,{\rm GeV}$.
\label{fig:photondistribution}
 }
 \end{center}
 \end{figure}

The meson decay therefore yields a distribution of real photons depending on the energy $E_{\gamma}$ and the angle $\theta_{\gamma}$ with respect to the beam axis. Due to symmetry there is no dependence of this photon distribution with respect to rotations around the beam axis.
The distribution is shown in Fig.~\ref{fig:photondistribution} for proton beam energies of $400\,{\rm GeV}$ and $70\,{\rm GeV}$.

\bigskip

In the second step the photons produce ALPs via the Primakoff conversion on the nucleus. 
We use laboratory-frame coordinates, where the nucleus is at rest.
By a suitable rotation we can choose the photon to be moving in the $z$-direction. We then have, 
\begin{equation}
p_{\gamma}=\left(\begin{array}{c}
\omega\\
0\\
0\\
\omega
\end{array}
\right)
\qquad\qquad\qquad\qquad
p_{a}=\left(\begin{array}{c}
E_{a}\\
0\\
k_{a}\sin(\photontheta)\\
k_{a}\cos(\photontheta)
\end{array}\right).
\end{equation}
Note, however, that in this coordinate system the $z$-direction is slightly rotated compared to the beam axis, with the angle arising from the angle of the meson production and subsequent decay into the desired photon. In the Monte-Carlo we then rotate back to the system with the $z$-axis along the beam axis\footnote{In practice we do this by using a coordinate system as in~\cite{Dobrich:2015jyk} that allows for an angle with respect to the beam axis right from the beginning.}.

We use the approximate form of the cross section as in~\cite{Dobrich:2015jyk}\footnote{We have checked by comparing to more complete expressions in~\cite{Cadamuro:2010cz,Aloni:2019ruo} that the approximation to the cross 
section is excellent in the regime giving relevant contributions within the experimental acceptance.}. In this simple coordinate system the cross section, approximated for small $\vartheta$, reads,
\begin{equation}
 \frac{\rm{d}\sigma_{\rm P}}{\rm{d}\photontheta}= \frac{4\pi  \, \alpha Z^2 g^{2}_{a\gamma} \ \omega^8 \ \photontheta^3 \ |F_{em}(q)|^2}{(m^4_{a} + 4 \ \omega^4 \ \photontheta^2)^2} \ .
 \label{eq:primakoff}
\end{equation}
As in~\cite{Dobrich:2015jyk}
the electromagnetic form-factor $F_{em}$, is taken to be of the Helm form\footnote{At very low momentum transfer $q\lesssim 10\,{\rm keV}$ (depending on the target material) the electron shell also shields the charge, reducing the form factor. However, as discussed in~\cite{Dobrich:2015jyk} this region only gives a very small contribution to the signal which we neglect here.}  (cf.~\cite{Woods:1954zz}),
\begin{equation}
\label{eq:formfactor}
F_{em}(q^2) = \frac{3 \, j_1(\sqrt{q^2}\,R_1)}{\sqrt{q^2}\,R_1} \exp\left[-\frac{(\sqrt{q^2}\,s)^2}{2}\right] \; ,
\end{equation}
with $j_1$ the first spherical Bessel function of the first kind.
For the nuclear radius we use~\cite{Lewin:1995rx}
\begin{equation}
\label{radius}
R_1 = \sqrt{(1.23 \, A^{1/3}-0.6)^2+2.18}\,\, {\rm fm}\,\, .  
\end{equation}
To simplify the evaluation we set the form factor to zero for values $q \, R_1 \geq 4.49$, i.e. above the first zero of the Helm form factor (as in~\cite{Dobrich:2015jyk}). 
 
Let us also emphasize again that in the present case we are dealing with a distribution of real, on-shell photons. This is in contrast to the photon from proton (PFP) mode where we have an effective parton distribution of virtual photons.
Therefore, in this production mode we do not expect to be affected by the corrections to the equivalent photon approximation discussed in~\cite{Harland-Lang:2019zur}.

The cross section is then determined by folding the (probability) distribution of the photons from the mesons with the cross section for 
an individual photon\footnote{Let us note at this point that formally the PFP production and the production from decay photons is at different order in $\alpha$. Meson production is a strong process and therefore essentially independent of $\alpha$ whereas
``radiating'' a photon from the proton requires an extra electromagnetic interaction. 
This is accounted for by a factor of $\alpha$ in the relevant photon distribution functions for the proton (cf. also~\cite{Berlin:2018pwi}).}.

\begin{figure}[t]
\begin{center}
  \includegraphics[width=0.49\textwidth]{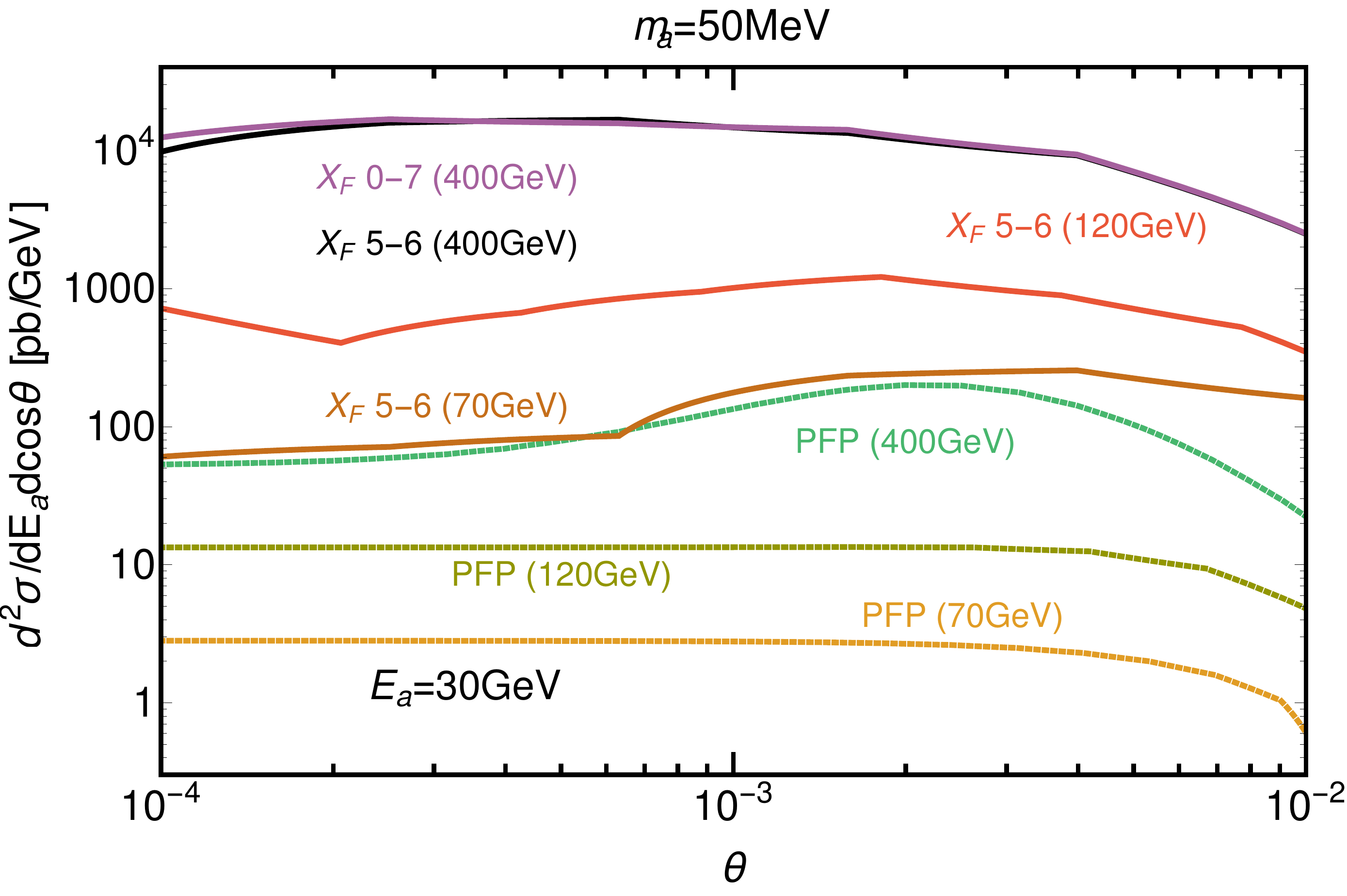} 
 \includegraphics[width=0.49\textwidth]{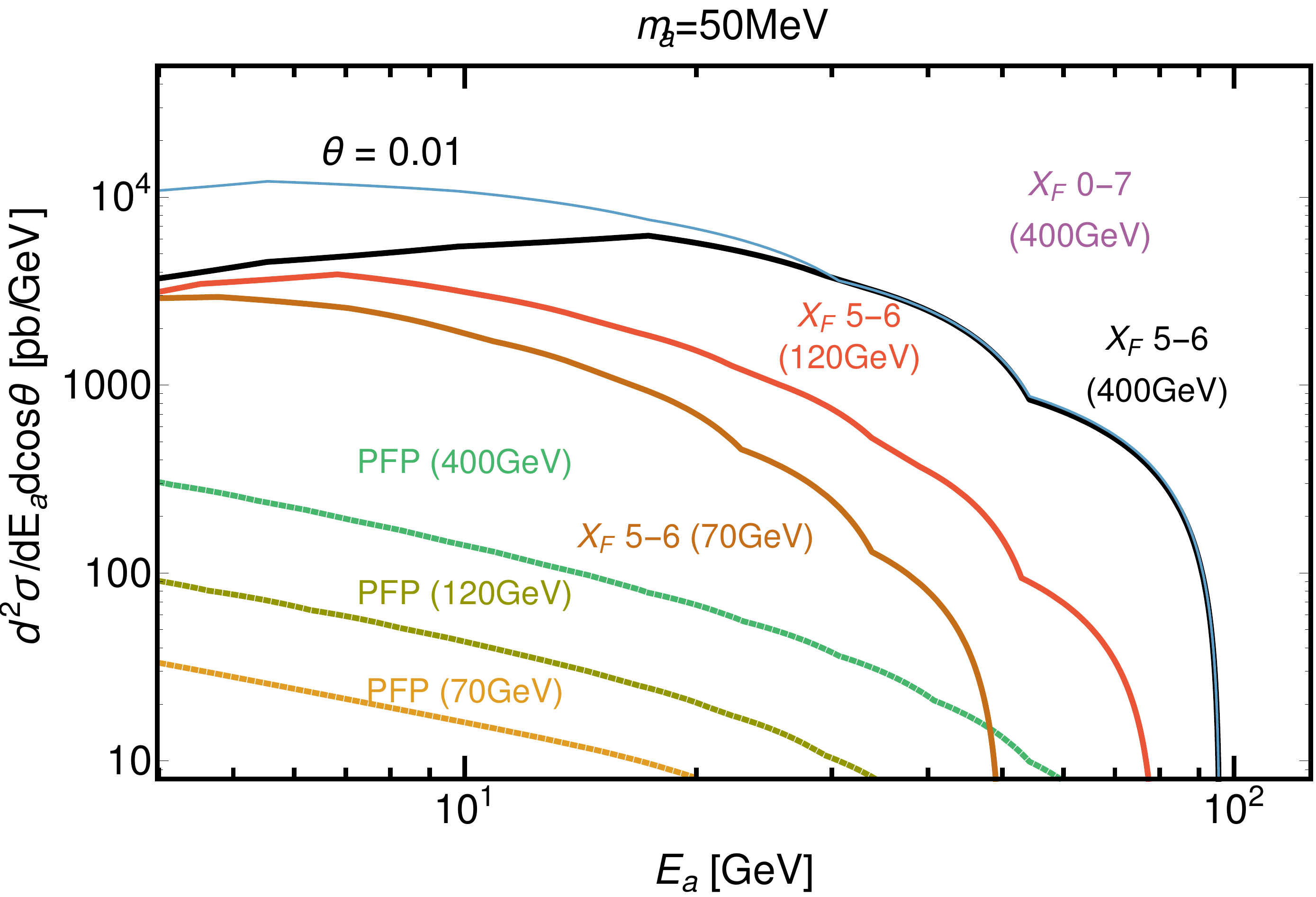} 
   \includegraphics[width=0.49\textwidth]{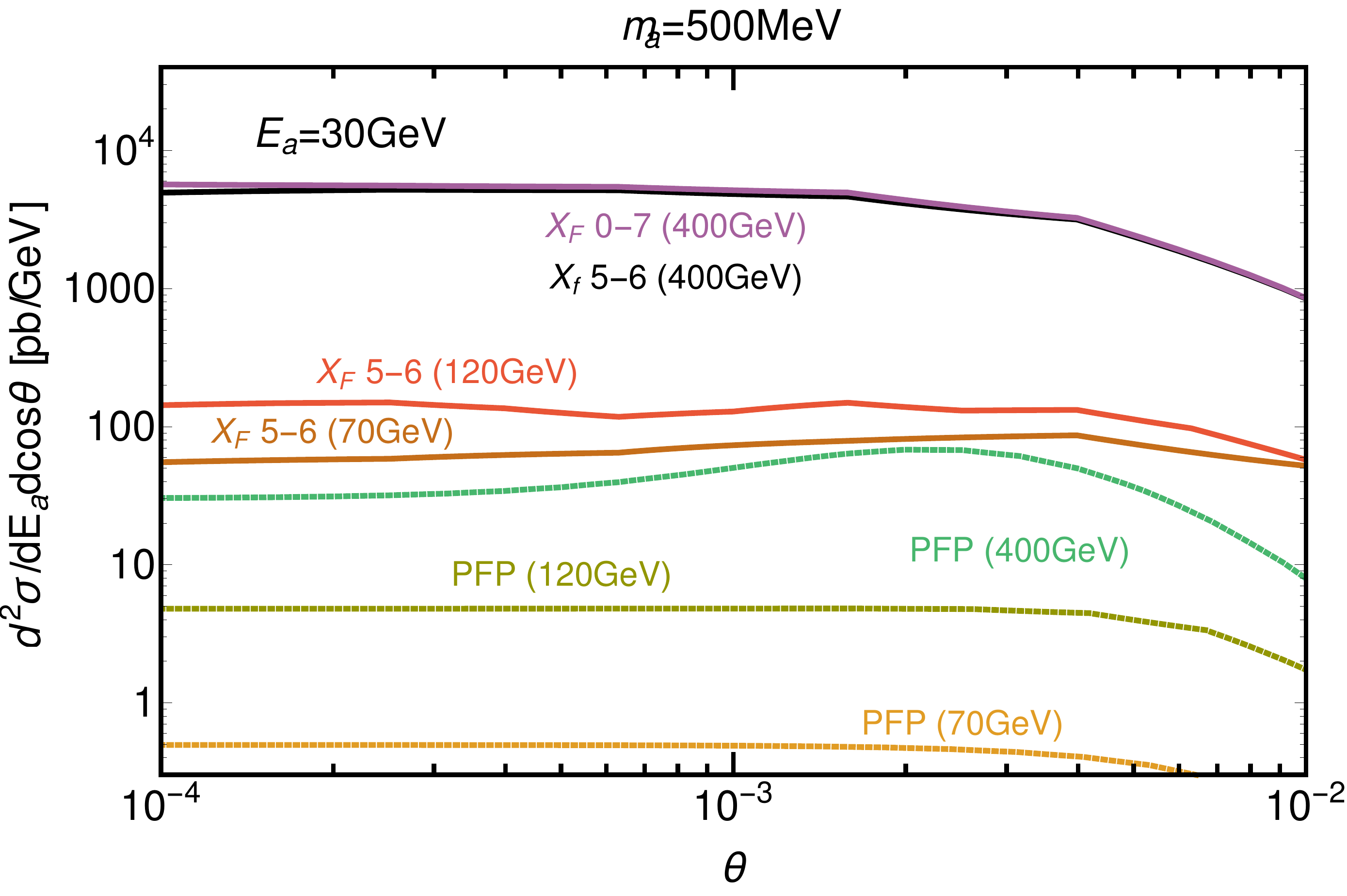} 
 \includegraphics[width=0.49\textwidth]{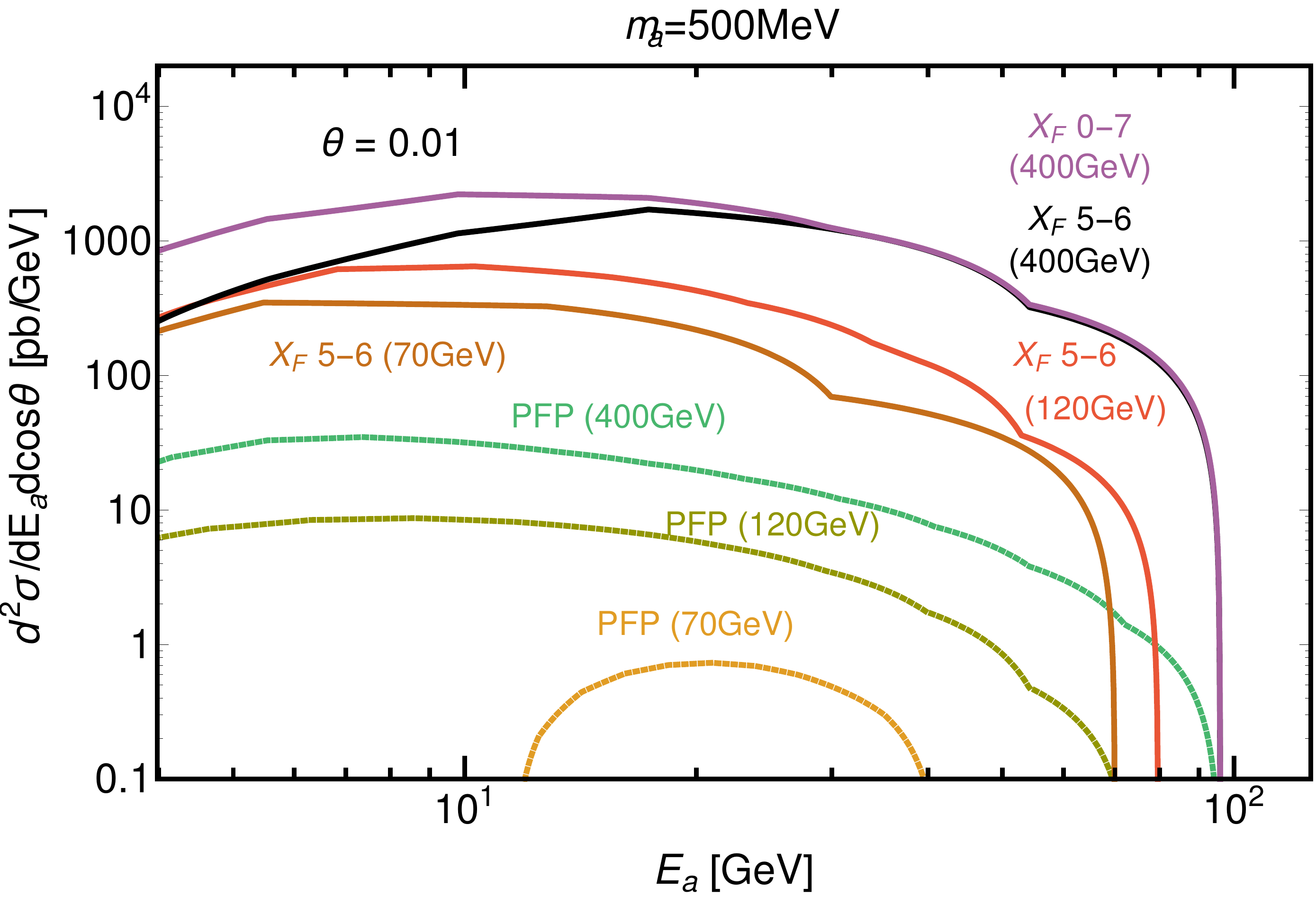} 
 \caption{\label{fig:reach_photon_distris}
 Differential ALP production cross-sections as a function of angle (left) for fixed ALP energy $E_a=30\,{\rm GeV}$
 and as a function of ALP energy for fixed angle $\theta=0.01$ (right).
Top and bottom correspond to an ALP mass  of $m_a=50\,{\rm MeV}$ and $m_a=500\,{\rm MeV}$, respectively.
The dotted lines denote the  PFP contributions to the ALP production 
for a $400\,{\rm GeV}$ (green), $120\,{\rm GeV}$ (olive) and  $70\,{\rm GeV}$ (gold) 
proton beam.
Solid lines denote the contribution of ALP production through decayed $\pi^{0}$s.
For $400\,{\rm GeV}$, we include examples of the combined contribution of all $X_F$ bins 0-7 or
the central part of $X_F$ bins 5-6 only, as defined in Table~\ref{tab:xfBins}.
These curves are shown at
a primary proton energy of $400\,{\rm GeV}$ in purple ($X_F$ bins 0-7) and black ($X_F$ bins 5-6), 
$120\,{\rm GeV}$ (red, $X_F$ bins 5-6) and $70\,{\rm GeV}$ (brown, $X_F$ bins 5-6).
The chosen target materials are copper at $400\,{\rm GeV}$, and iron at $120\,{\rm GeV}$ and 
$70\,{\rm GeV}$.
The slight kinks in the lines arise from the logarithmic binning used for the presented curves. For the sensitivity calculation, a finer binning has been used.
}
 \end{center}
\end{figure}

As an example we show in Fig.~\ref{fig:reach_photon_distris} the resulting cross sections for several energies and target materials relevant for the analysis in the next section (solid lines).
This is then compared to the cross section for the PFP mode shown as dashed lines.
For all considered energies $70\,{\rm GeV}$, $120\,{\rm GeV}$ and $400\,{\rm GeV}$, the PFP mode is sub-dominant. Also,
for all energies, production from mesons is particularly favored since the spectrum is also considerably harder.
The harder spectrum is useful for detecting ALPs with relatively large couplings since the higher $\gamma$-factor allows ALPs to decay outside dump and shielding regions.

\bigskip
For a better understanding of the behavior shown in Fig.~\ref{fig:reach_photon_distris} we note that the Primakoff cross section, Eq.~\eqref{eq:primakoff}, is peaked at small angles. More precisely, it is peaked at angles $\vartheta\sim m_{a}^2/\omega^2$. 
For sufficiently small masses and high energies this angle is very small. The ALP then has the same angle with respect to the beam axis as the incoming decay photon. Therefore, to a large degree the energy and angular dependence is dominated by that of the initial photon distribution shown in Fig.~\ref{fig:photondistribution}. 
However, this is only an approximate statement since the cross section~\eqref{eq:primakoff} also has a significant tail towards larger angles where it is ultimately cut off by the form factor.

To get a better quantitative feeling for the involved scattering angles let us note that the radii, according to Eq.~\eqref{radius}, are in the range $R_{1}\sim (2-5)\,{\rm fm}$ for the elements we are considering. With $|q|\sim E_{a}\vartheta$ for small masses, and the form factor setting in at $|q|R_{1}\sim 1$ we find that the tail is typically starting to be cut off at angles $\vartheta\sim (0.002-0.004)\,(20\,{\rm GeV}/E_{a})$. For energies $\gtrsim 20\,{\rm GeV}$, 
initial photon angles within a few milliradians of the desired ALP angle contribute. Stated differently, for energies $E_{a}\gtrsim 20\,{\rm GeV}$ and angles $\gtrsim10\,$milliradians, 
the ALP angle is dominated by the angle of the decay photon given in Fig.~\ref{fig:photondistribution}. 

\bigskip 

Fig.~\ref{fig:reach_photon_distris} also demonstrates our procedure to estimate the uncertainty in our production rate. For $400\,{\rm GeV}$ we show two lines for the meson production. One (purple) in which we include all mesons in the full kinematic region and a more conservative estimate (black) where we include only the part of the meson spectrum that is most trustworthy. To be more precise we define two regions in $X_F$. One covering all the bins, 0-7, in Tab.~\ref{tab:xfBins}, whereas the other, more conservative one includes only the bins 5-6 where we find better agreement between Monte Carlo and data in our validation procedure.
While this is very conservative, it does not fully appreciate the degree to which we underestimate the cross section. Indeed bin 6 gives a significant contribution to the cross section. From Fig.~\ref{fig:LEBC_vs_PYTHIA_xf} we can see that in this region the simulated cross section underestimates the cross section by a factor which can easily be $\sim 2$. Our conservative estimate therefore underestimates the cross section by perhaps ${\rm few}\times 10\%$. Validated spectra improving in this region could therefore allow for a significantly increased sensitivity.

\bigskip
In the third step we then take into account all the relevant experimental cuts,
\begin{itemize}
\item{} The ALP decay has to happen outside the target and in the decay volume.\\ Crucially this gives the exponential dependence on the decay length, \linebreak \mbox{$\exp\left(-\frac{d}{\ell_{\rm ALP}}\right) - \exp\left(-\frac{d+l}{\ell_{\rm ALP}}\right)$,} where $\ell_{\rm ALP}$ is the decay length, $d$ the distance from the target to the decay volume and $l$ the length of the decay volume.
\item{} The decay photons of the ALP have to reach the detector with a suitable minimum energy and other criteria are required for the photons to be detected.
\end{itemize}
We will describe the relevant details and approximations when we discuss the individual experiments.
After all those cuts we obtain the fiducial cross section,
\begin{equation}
\sigma_{f}={\rm fiducial\,\,cross\,\,section\,\,after\,\,all\,\,cuts}.
\end{equation}

Before we take the final step, let us also note that we have employed the simplification that both the meson production and the photon conversion into ALPs happen at the beginning of the target. This neglects the finite distance traveled before the proton interacts inside the target as well as the additional distance traveled by the photon before it is converted into an ALP. The typical distances are of the order of the proton and photon radiation length, respectively. As both are of the order of cm in the relevant energy range, this should be a minor effect compared to the target sized of the order of m for the experiments we will consider in Section~\ref{sec:sensi}.
For experiments with targets in the cm range this would have to be taken into account.

\bigskip
In the final step we now need to compare the fiducial cross section for the detectable ALP production from a photon with the cross section for the photon to be absorbed in the target material. We then get for the total number of events,
\begin{equation}
N=N_{\gamma}\frac{\sigma_{f}}{\sigma_{\gamma, {\rm target}}},
\end{equation}
where $N_{\gamma}$ is the total number of photons produced in the meson decays. The total cross section $\sigma_{\gamma, {\rm target}}$ for photons to be absorbed in the target material can be determined from the radiation length $\ell^{\rm rad}_{\rm target}$,
\begin{equation}
\sigma_{\gamma, {\rm target}}=\frac{m_{N}}{\rho_{\rm target}\ell^{\rm rad}_{\rm target}},
\end{equation}
where $m_{N}$ is the mass of the nucleus.

We stress that this is different from the PFP mode, where the relevant cross section would be the cross section 
for proton-nucleus interactions.

For the photon cross section we use values for the radiation length from~\cite{gammaxs}, whereas for the proton cross section we employ $\sigma_{pN} \simeq 53\:\text{mb} \times A^{0.77}$ from~\cite{Carvalho:2003pza},
where $A$ is the mass number of the nucleus. 
For example in the case of copper the photon cross section for energies $\gtrsim 2\,{\rm GeV}$ is about $6\,{\rm barn}$
for copper, $5.1\,{\rm barn}$ for iron and $12.7\,{\rm barn}$ for molybdenum. This is in contrast to the proton cross section which is only $1.3\,{\rm barn}$, $1.2\,{\rm barn}$ and $1.8\,{\rm barn}$, respectively.

While the photon absorption cross section is significantly larger we find that in many cases this is more than compensated by a number of other factors. In particular we produce on average more than 1 meson per proton and each meson gives two photons. Importantly, as we have seen in Fig.~\ref{fig:reach_photon_distris}, the spectrum of ALPs from these photons is also harder than the one from the PFP mode, which is advantageous for the detection in particular at larger ALP masses.

\section{Updated sensitivities for fixed-target experiments \label{sec:sensi}}

In the following, we will first update the exclusion contours for the
past fixed target experiments CHARM~\cite{Bergsma:1985qz} and NuCal with ALPs
produced from the decay of $\pi^0$.
We then project sensitivities for the existing NA62~\cite{NA62:2017rwk} and SeaQuest~\cite{Berlin:2018pwi} set-ups.
Finally, we give projections for the proposed SHiP~\cite{Alekhin:2015byh,Anelli:2015pba} facility.

\subsection{Past experiments: NuCal and CHARM update}

For CHARM~\cite{Bergsma:1985qz}, we make use of the following parameters: 
The detector was
located at a distance $d=480\,\mathrm{m}$ away 
from the proton dump, and the protons were dumped into a copper target.
The detector (for CHARM this is
identical to the decay volume) was $l=35\,\mathrm{m}$ in length and 
$3\times3\,\mathrm{m}$ 
in transverse dimensions. The detector is off-set transversally by
$5\,\mathrm{m}$ from the beam axis and this is accounted for in the MC. 
According to Ref.~\cite{Bergsma:1985qz}, CHARM was sensitive to events with 
a single electromagnetic shower in acceptance. CHARM quotes a signal 
acceptance of 51\%, which we include in our estimate. The number of protons on
target (POT) is $2.4 \times 10^{18}$.

NuCal~\cite{Blumlein:1990ay}, made use of the U70 proton beam facility with a beam energy of
`only' $70\:\text{GeV}$.
However, NuCal profits from a comparably small
distance between target and detector of only $d= 64\:\text{m}$ and
a detector length of $23\:\text{m}$. We adapt the analysis strategy of~\cite{Blumlein:2013cua}, 
and require a minimum ALP energy of $E_a > 10\:\text{GeV}$ and at least one photon detected.
In this way, the acceptance is approximately constant and equal to $70\%$. 
The detector has a radius of $1.3\:\text{m}$.
In a dataset of $N_\text{pot} = 1.7 \cdot 10^{18}$ protons on an iron target, NuCal observed
1 event compared to a background expectation of 0.3 events. At 90\% confidence level,
we can therefore exclude any point in the parameter space predicting more than 3.6 events.

\bigskip

The changes in the limits from CHARM and Nucal when including ALPs produced 
from decayed $\pi^0$s, can be understood from Figure~\ref{fig:reach_photon_distris}.

For NuCal, where the beam energy is 70~GeV, the plot in the right-hand side of the
figure illustrates that, including the $\pi^0$ yield
appreciably changes the existing limits. This can be
seen in Figure~\ref{fig:new_lims} (l.h.s.), where the brown dashed line shows the NuCal
limits in PFP mode only, while the yellow region is the reach
considering the added yield for NuCal from PFP and a conservative (bins 5--6) $X_F$ range from decayed mesons. 
However, the position of the upper part of the exclusion contour (at large couplings)
is mostly determined by the experiment's geometry. Thus, even the revised
NuCal limit does not drastically alter the untested parameter space that should be probed by new experiments,
discussed in the next section. 

Also for CHARM, including the ALP production channel via $\pi^0$s improves the 
sensitivity with respect to the CHARM reach from PFP considerably\footnote{A recent study
\cite{Dobrich:2018jyi} on ALPs coupled to fermions
is another example of the importance of carefully recasting results from past experiments,
particularly CHARM,  using PYTHIA and MC simulations to appreciate their full impact.}.
Figure~\ref{fig:new_lims} (l.h.s.) shows a zoomed version of the existing limits with updated CHARM contour 
(solid red) compared to the previous curve (dashed magenta).
However, the CHARM limits lie still within the limits of E137 and the new NuCal limit.
Thus
we summarize that the experimental future landscape, does not change significantly even after the CHARM and NuCal updates.
However, as we will see is the next subsection, the inclusion of the meson contribution, drastically changes
the {\it prospect} sensitivity for forthcoming searches.

\subsection{Current and future set-ups}

The NA62 experiment~\cite{NA62:2017rwk} has been built to achieve a precise measurement of the ultra-rare 
decay $K^+\rightarrow \pi^+ \nu \bar{\nu}$. 
Besides its main goal,
NA62 has a rich program to search for exotic particles, including long-lived particles that can be produced 
in the up-stream copper beam collimator, 
into which the primary SPS proton-beam is 
fully (in dump-mode) or partially (in standard, parasitic data-taking) dumped (see, e.g. \cite{Dobrich:2018ezn} for 
more details).

To be sensitive to a fully neutral final state, NA62 has to be run in beam-dump-mode.
In our MC, we model NA62 using the following parameters: The distance between the beam-defining collimator (used to dump the beam) and the start of the fiducial 
volume is $d = 82\,\mathrm{m}$ and the vacuum decay region before the 
Liquid Krypton Calorimeter (LKr) is $l=135\,\mathrm{m}$ long. 
In addition, we require the following acceptance conditions: 
Both photons produced in the ALP decay need to be detected at a minimum mutual distance of $10\,{\rm cm}$ in the LKr.
Moreover, these photons need to be $15\,{\rm cm}$ away from the LKr central hole and their combined energy needs to be above $3\,{\rm GeV}$. 
The target material for NA62 is copper and we show NA62 prospects for two different choices of POT.

\begin{figure}[]
\begin{center}
  \includegraphics[width=0.46\textwidth]{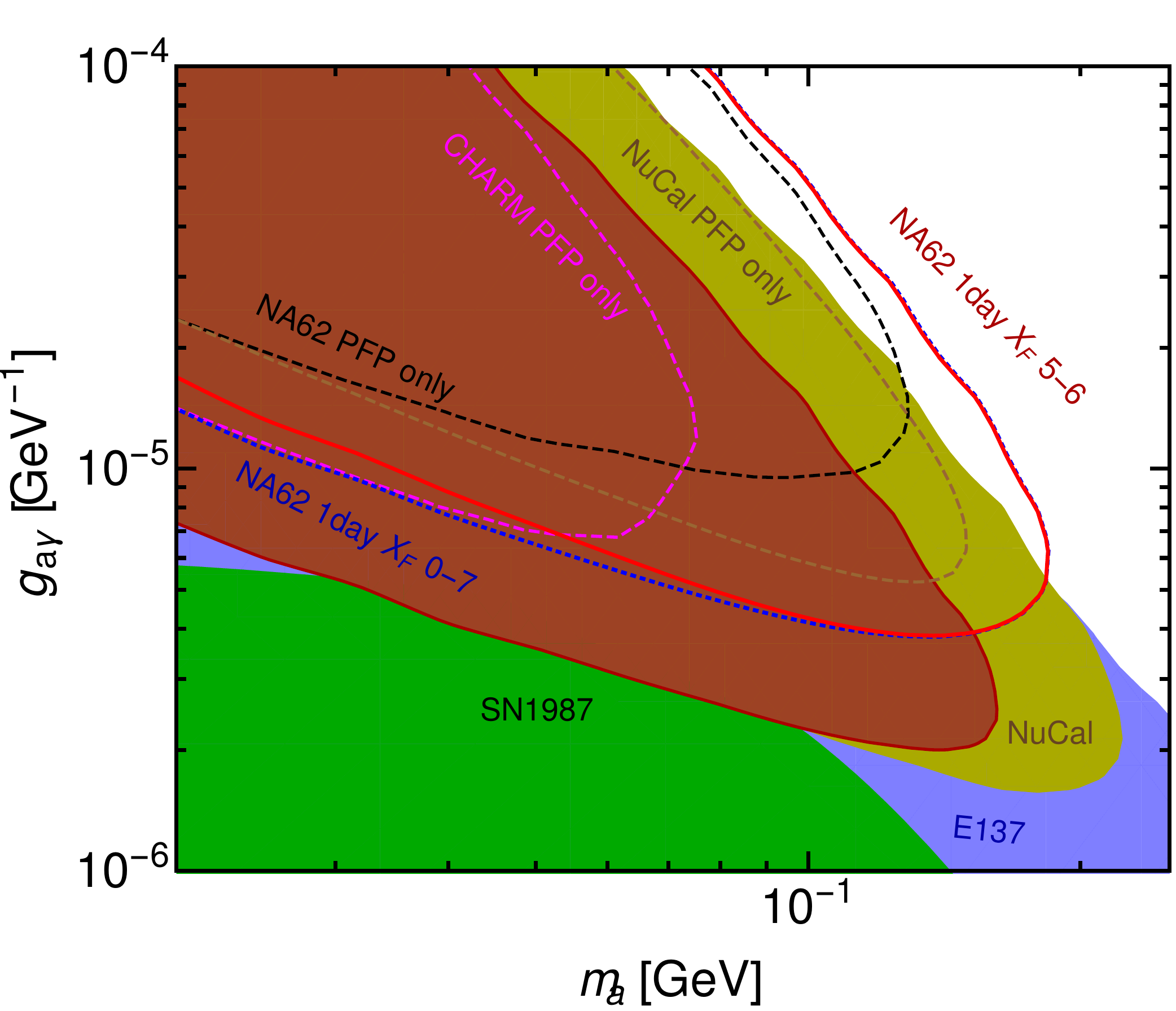}\hspace*{1cm}
 \includegraphics[width=0.46\textwidth]{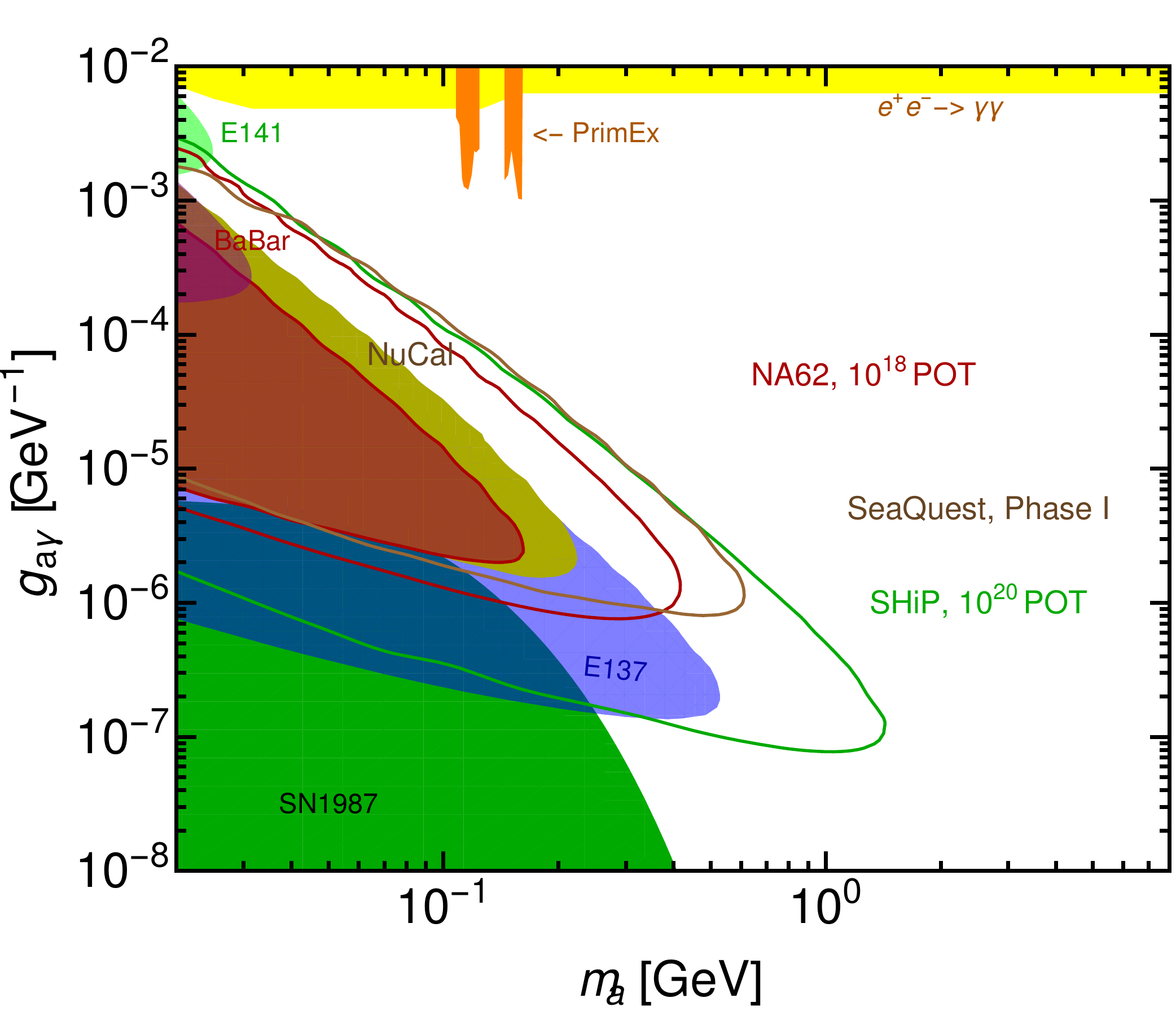} 
 \caption{\label{fig:new_lims}
Both plots: filled areas: 90\%-CL excluded regions from past experiments (cf.~\cite{Dolan:2017osp,Aloni:2019ruo}); 
contours: projected 90\%-CL exclusion capability at present or future experiments.
Left: Excerpt of the limit plot as in Fig~\ref{fig:ALPs_status} but with updated CHARM region (brown):
including PFP and $\pi^0$
contributions from $X_F$ bins 5--6, magenta dashed contour: PFP alone).
In addition we show
three projections for NA62 at $1.3 \times 10^{16}$ POT (1 day):
Black dashed: PFP alone, red solid and blue dotted: Yield from $\pi^0$
in using $X_F$ bins 5--6 and 0--7, respectively.
Right: projections for NA62 at $1 \times 10^{18}$ POT,
as well as 
SHiP and SeaQuest (Phase I) at $1 \times 10^{20}$ POT and $1.44 \times 10^{18}$ POT, respectively.
We have checked that the MC statistical uncertainty is negligible compared to the uncertainties 
of the PYTHIA/data agreement (see Section~\ref{sec:yield}).
}
\end{center}
\end{figure}

In Figure \ref{fig:new_lims} (l.h.s.) we show the sensitivity for NA62
at $1.3 \times 10^{16}$ POT (corresponding to a one-day run), 
using (purple dashed) PFP only,
using the full $X_F$ range (bins 0--7, blue dotted) and the central $X_F$ bins 5--6 (red), respectively.
We also show the sensitivity for $1.3 \times 10^{16}$ POT PFP to facilitate comparison with previous 
results~\cite{Dobrich:2015jyk}.
As shown, the contribution from the decay of $\pi^0$'s
dominates the sensitivity prospect for a $400\,{\rm GeV}$ proton beam
and constitutes one of the most relevant results of our study.
Regarding the question, whether all or only a conservative number
of $X_F$ bins should be chosen, we observe that
the difference between the reach in both situations is small
but visible at ALP masses of few tens of MeV and small couplings.
This can be understood by looking at Figure~\ref{fig:reach_photon_distris}.
The central $\pi^0$ production corresponding to $X_F$ bin 4 contributes to low photon momenta and correspondingly low ALP momenta. As we can see in Fig.~\ref{fig:new_lims} this is more relevant for the 
sensitivity at low masses and couplings. The forward contribution to $\pi^0$ production corresponding to bin 7, is not relevant in increasing the cross section at high photon energies.
Not including $X_F$ bins other than 5 and 6 has a minor impact on our projections
and thus for all following computations we adopt this conservative condition.

In Figure~\ref{fig:new_lims} (r.h.s) we show again the prospect for NA62, this time
however at $10^{18}$~POT (corresponding to a few months of data taking) while summing up contribution
from $\pi^0$ production and PFP. 
Comparing this to the prospects shown, e.g. in~\cite{Beacham:2019nyx,Dobrich:2018ezn}
underlines the importance of including the yield of ALPs produced by $\pi^0$s  in these estimates.

The sensitivity reach for SeaQuest to ALPs produced as a result of secondary $\pi^0$ decays 
was estimated previously in~\cite{Berlin:2018pwi}. As outlined in~\cite{Berlin:2018pwi}, the
sensitivity to a di-$\gamma$
final state requires the installation of an ECAL, potentially adapted from the
PHENIX detector at BNL. Our analysis improves the study put forward
in~\cite{Berlin:2018pwi} in several key regards. Firstly, as outlined in Sect.~\ref{sec:yield}, we have validated
the $\pi^0$ differential cross sections obtained in PYTHIA against experimental literature in a wide energy range.
Secondly, as for all considered set-ups, we have implemented
a full Monte Carlo of ALP production and decay according to the geometry laid out in~\cite{Berlin:2018pwi}.
Lastly, we consider also the sub-dominant PFP contribution for ALP production in our estimate.

We use the geometric setup described in~\cite{Berlin:2018pwi}.
Also, as in~\cite{Berlin:2018pwi}, we assume the need for 10 signal events to detect a signal beyond the background fluctuations.
The calorimeter is placed between tracking stations 3 and 4 (at $\sim 18.5\,{\rm m}$ downstream of the target)
We use 
a fiducial volume that has a length of $1\,{\rm m}$, in between meters 7 and 8 of the experiment,
and a geometric acceptance of $2\times 2\,{\rm m}$ in transverse directions.
The target material of SeaQuest is iron and we assume the phase-I statistics of $1.44 \times 10^{18}$ POT.
Similarly to other set-ups, we require both photons to be detected at a minimum energy of $1\,{\rm GeV}$ each, a total energy of at least $3\,{\rm GeV}$
and
a minimum mutual distance of 10 cm
to avoid shower overlap given the photon shower Moliere radius.

The resulting prospects for SeaQuest are shown as the brown curve on the r.h.s. of Figure~\ref{fig:new_lims}. 
Compared to the estimates of~ \cite{Berlin:2018pwi}, the expected sensitivity
covers a somewhat larger area of parameter space.

Finally, we model the prospects for detection of ALPs in the SHiP~\cite{Alekhin:2015byh,Anelli:2015pba} calorimeter
as follows\footnote{The prospect geometry for SHiP has changed since the publication of
\cite{Dobrich:2015jyk} and we follow the layout of~\cite{shipCDS} for our estimate}:
The fiducial region is taken to be $45\,\mathrm{m}$ downstream of the production point. 
The calorimeter is positioned at $50\,\mathrm{m}$ after the beginning of the decay volume.
We ask both photons to be in an acceptance area of 5 $\times $10~m$^2$. Both photons should have a minimum energy
of 1 GeV and a combined energy of 3 GeV and be at least $10\,{\rm cm}$ apart.
The result can be seen in Fig~\ref{fig:new_lims} r.h.s as green curve. Target material is molybdenum
and the POT are $10^{20}$.
Note that the envisaged SHiP calorimeter~\cite{Bonivento:2018eqn} has the potential 
of reconstructing the photon direction, thereby allowing ALP mass reconstruction.
Compared to the results shown in~\cite{Beacham:2019nyx} the mass reach increases considerably,
from $\sim1\,{\rm GeV}$ to $\sim 1.5\,{\rm GeV}$.

\section{Conclusions}
\label{conclusions}
Proton beam dump experiments are a popular and versatile tool to explore the dark sector in the MeV to GeV range that may be connected to a number of open problems in particle physics, notably Dark Matter.
Important examples of current and near future experiments are NA62 (in beam dump mode), SeaQuest, and 
SHiP.

New particles can be produced in primary interactions of the proton with the target material but also in the decay of secondary mesons.
However, further important contributions to the production can arise from the interaction of secondary or even tertiary particles with the target material. While these production mechanisms have been noted and even occasionally used~\cite{Berlin:2018pwi,Feng:2018pew} they are still somewhat under-appreciated and many sensitivity calculations do not take them into account.

In this paper we have performed a detailed investigation of axion-like particle (ALP) production. More precisely we discussed the production of ALPs from the following process: Protons interact with the target nucleus and produce neutral mesons.
The mesons (mostly $\pi^{0}$) decay into two $\gamma$ which subsequently can interact with another target nucleus to produce an ALP via the Primakoff process.
We show that this gives a significant contribution to the production of ALPs which is also kinematically well suited for detection in typical experimental setups. Indeed, for experiments with high beam energies such as NA62 or SHiP, this is the dominant contribution in the region of interest and significantly extends the mass reach, e.g. by a factor of $\sim 1.5$ in the case of the SHiP experiment. 

A crucial input for the calculation of the production with secondary or even higher order particles are the spectra of these particles inside the target. In particular for mesons theoretical predictions are challenging due to the non-perturbative nature of the meson production processes. We have therefore validated our simulation results from PYTHIA 8.2 against a variety of measurements, thereby giving an estimate of the reliability of the simulation results and the impact this has on the sensitivity calculation for the experiments. 
We find that the impact of the uncertainty is moderate despite a relatively large uncertainty of our generated meson spectra in some regions of phase space.
Further, very desirable improvement could come from two directions. First of all, our simulations only include
mesons produced from the primary proton beam but, also for meson production, secondary interactions may play a sizable role. Including these secondaries will be an important next step. Second, the discrepancy of data and MC should be clarified by an extended study of existing data or, if needed, new measurements.

All in all, interactions of secondary particles in the beam dump are a powerful additional production mode for new very weakly coupled particles. Further studies that go beyond the example presented in this work are needed and in preparation.

\acknowledgments

We would like to thank M.~Pospelov for giving the nudge that started this work as well as P.~DeNiverville and 
F.~Kahlhoefer for very useful
discussions on the presented physics. We thank R.~Wanke for clarifications on the SHiP prospectus
setup. TS would like to thank P. Di Nezza for useful discussions on the validation of the MC simulation.
BD acknowledges support through the European 
Research Council (ERC) under grant ERC-2018-StG-802836 (AxScale).

\appendix
\section{Erratum}
\label{app:erratum}

\begin{figure}[h]
\begin{center}
  \includegraphics[width=0.45\textwidth]{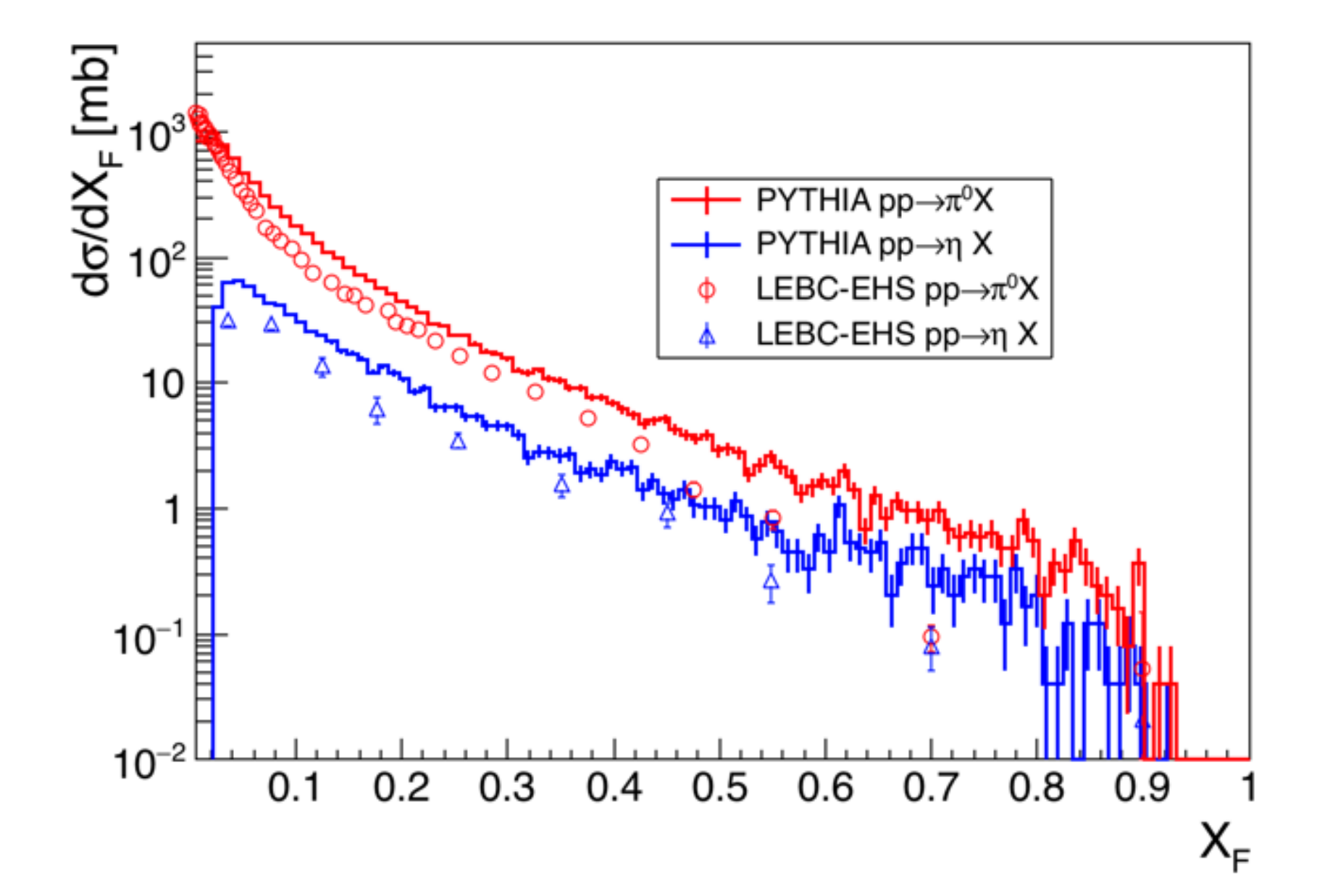}
  \includegraphics[width=0.45\textwidth]{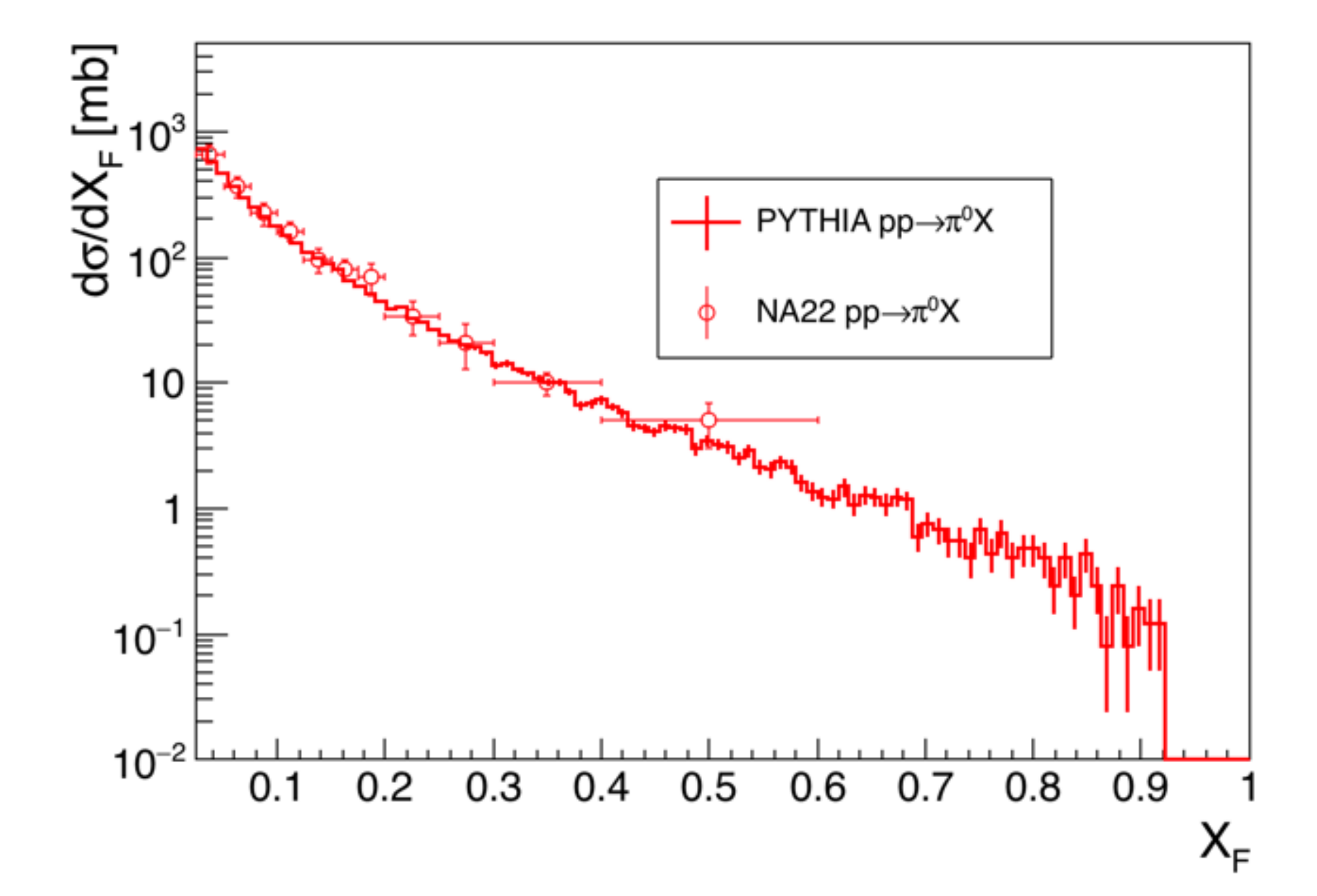}
  \hspace*{1cm}
 \caption{\label{fig:new_fig3}
Data-MC comparison of the differential cross-section
of neutral pions and eta mesons produced from 400~GeV (250~GeV) protons, to replace Fig~3 (the right panel of Fig~6) in JHEP~05~(2019)~213. 
}
\end{center}

\end{figure}

\begin{figure}[h]
\begin{center}
  \includegraphics[width=0.45\textwidth]{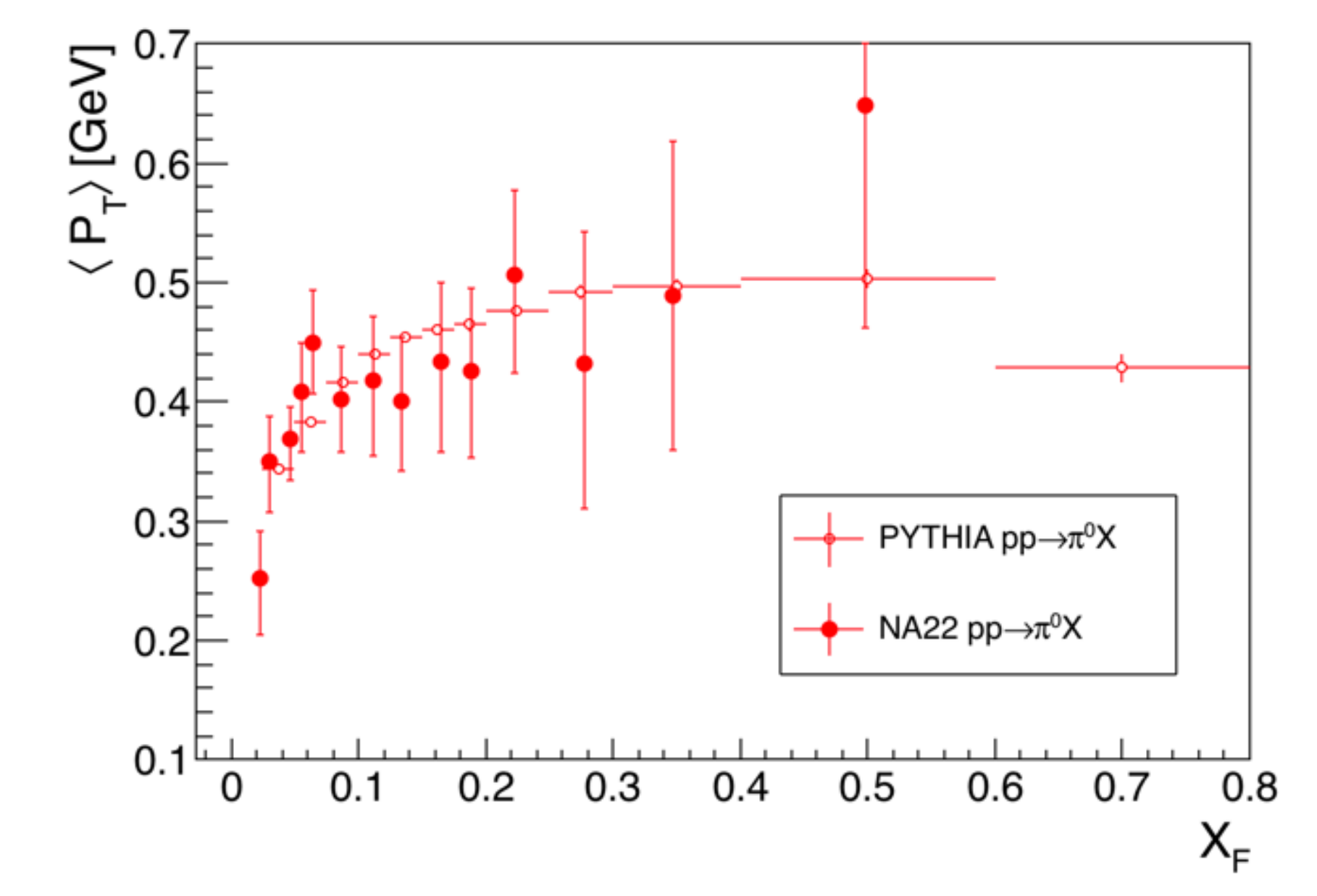}
  \includegraphics[width=0.45\textwidth]{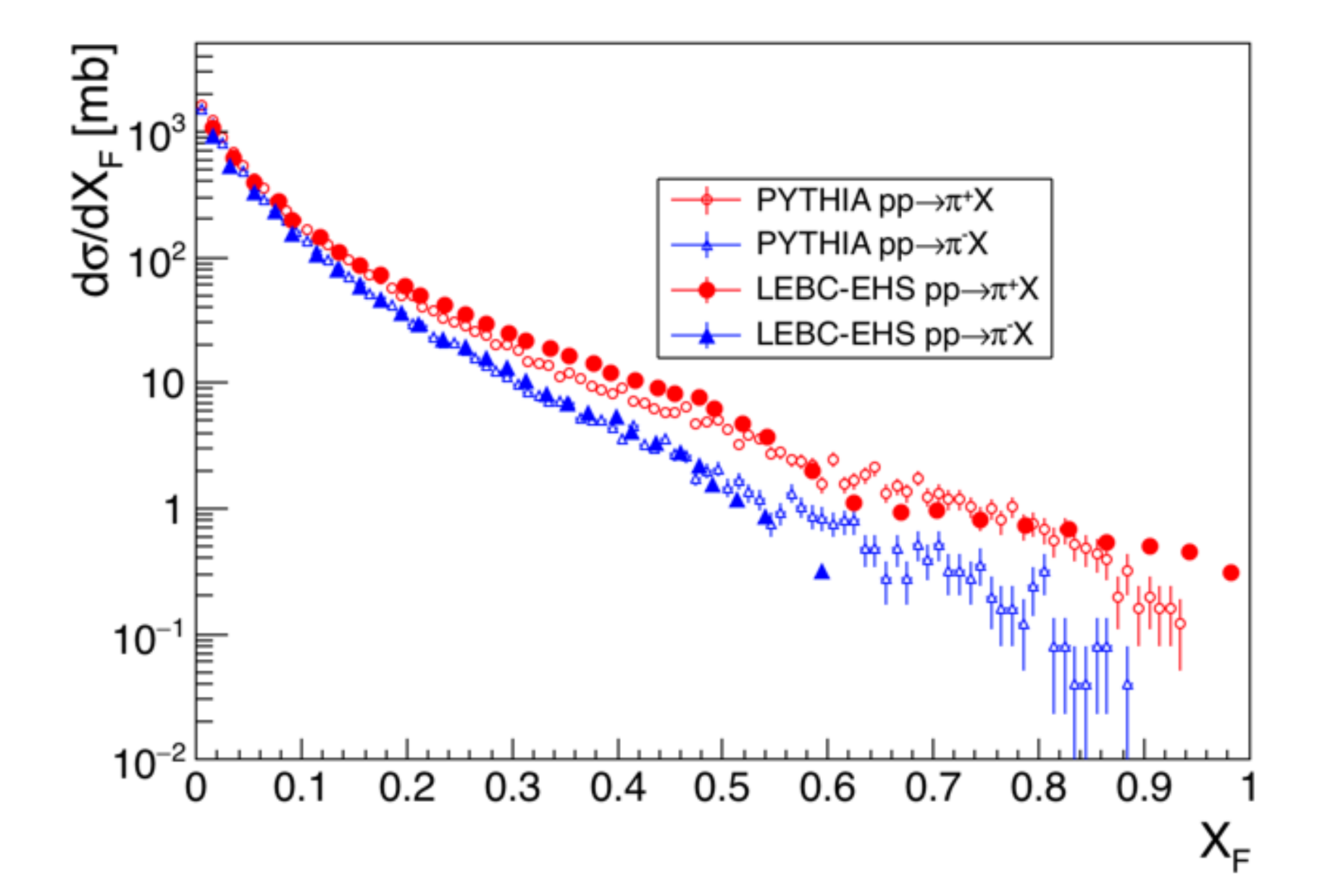}
  \hspace*{1cm}
 \caption{\label{fig:ptvsxf_bugfix}
Left: Data-MC comparison for the average transverse momentum as a function of the Feynan $X_F$ for a beam momentum of 250~GeV, to replace Fig~7 in JHEP~05~(2019)~213. 
Right: Data-MC comparison of the differential cross-section of charged pions produced from 400~GeV protons.
}
\end{center}
\end{figure}

 In \cite{Dobrich:2019dxc}, the differential cross section for inclusive $\pi^0$ production in proton-proton interactions is shown as a function the
 Feynman variable $X_F = P_Z/P_Z(\mathrm{max})\sim 2P^\ast_Z/\sqrt{s}$.
 The evaluation of $X_F$ after PYTHIA MC simulation was lacking a factor 2, affecting the display of the results for 250 and 400~GeV beam energy, in the right panel of Fig.~6 and in Fig.~3 of~\cite{Dobrich:2019dxc}, respectively. After correcting for this mistake, the new distributions are shown in Fig~\ref{fig:new_fig3}.
 Contrary to the distribution shown in~\cite{Dobrich:2019dxc} the MC now populates $X_F$ up to large values and a 
 segmentation of the variable at high $X_F$ to quantify the data-vs-MC
 agreement as done in Table 1 of~\cite{Dobrich:2019dxc} seems inappropriate. The agreement with the data is better for 250~GeV protons from the NA22 experiment~\cite{Adamus:1986ta} than for 400~GeV protons from the LEBC-EHS experiment~\cite{AguilarBenitez:1991yy}.
 The correlation between average transverse momentum and $X_F$ in the left panel of Fig.~\ref{fig:ptvsxf_bugfix} shows a good agreement between data and MC at 250~GeV, too. 
 

 At 400~GeV beam energy, the PYTHIA predicted yield is higher than the data especially at large $X_F$. To further scrutinize this issue, we first present the same comparison for charged pions at 400~GeV beam energy: as shown in the right panel of Fig.~\ref{fig:ptvsxf_bugfix}, the data-MC comparison is significantly better than for $\pi^0$ mesons.

\begin{figure}[h]
\begin{center}
  \includegraphics[width=0.45\textwidth]{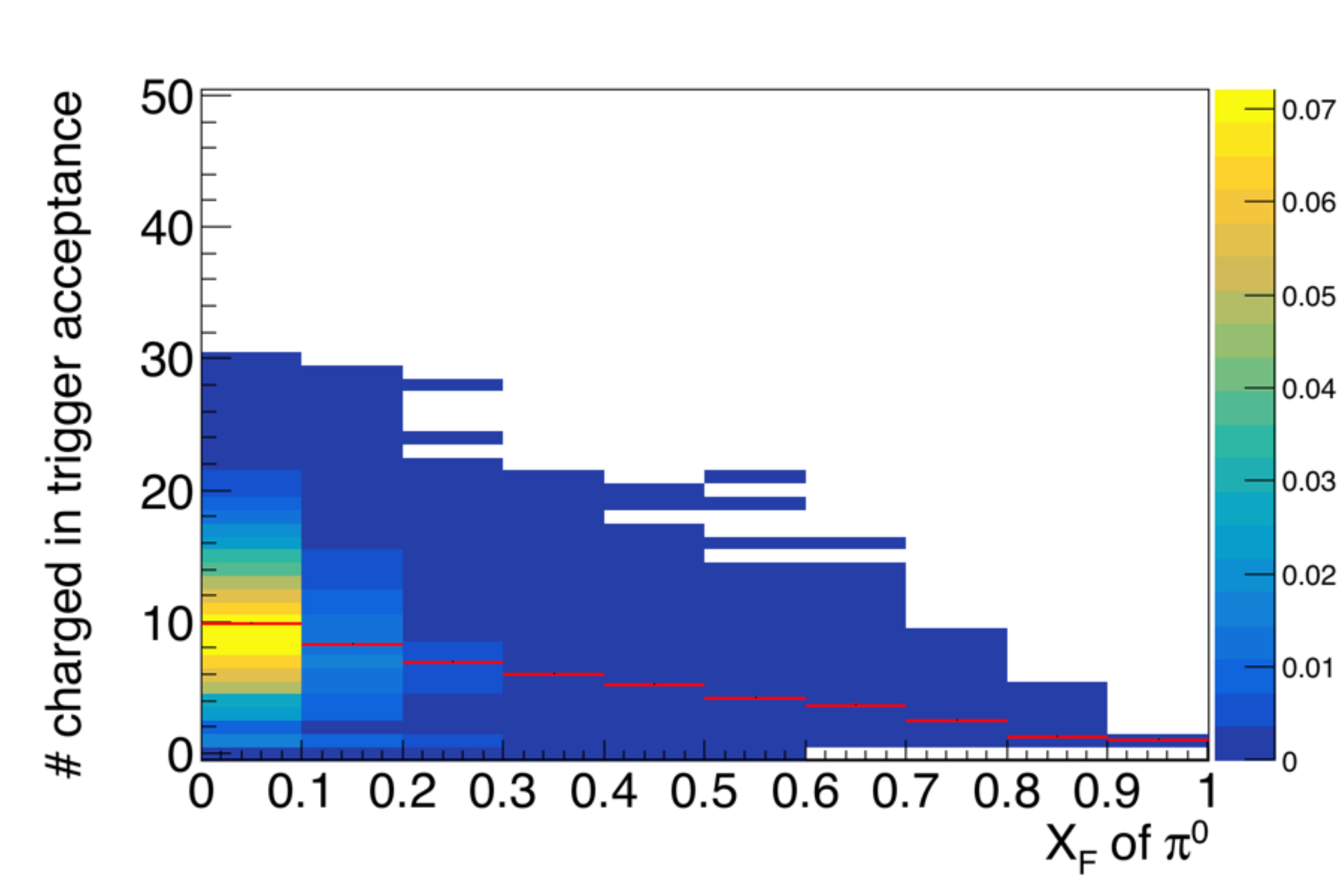}
  \includegraphics[width=0.45\textwidth]{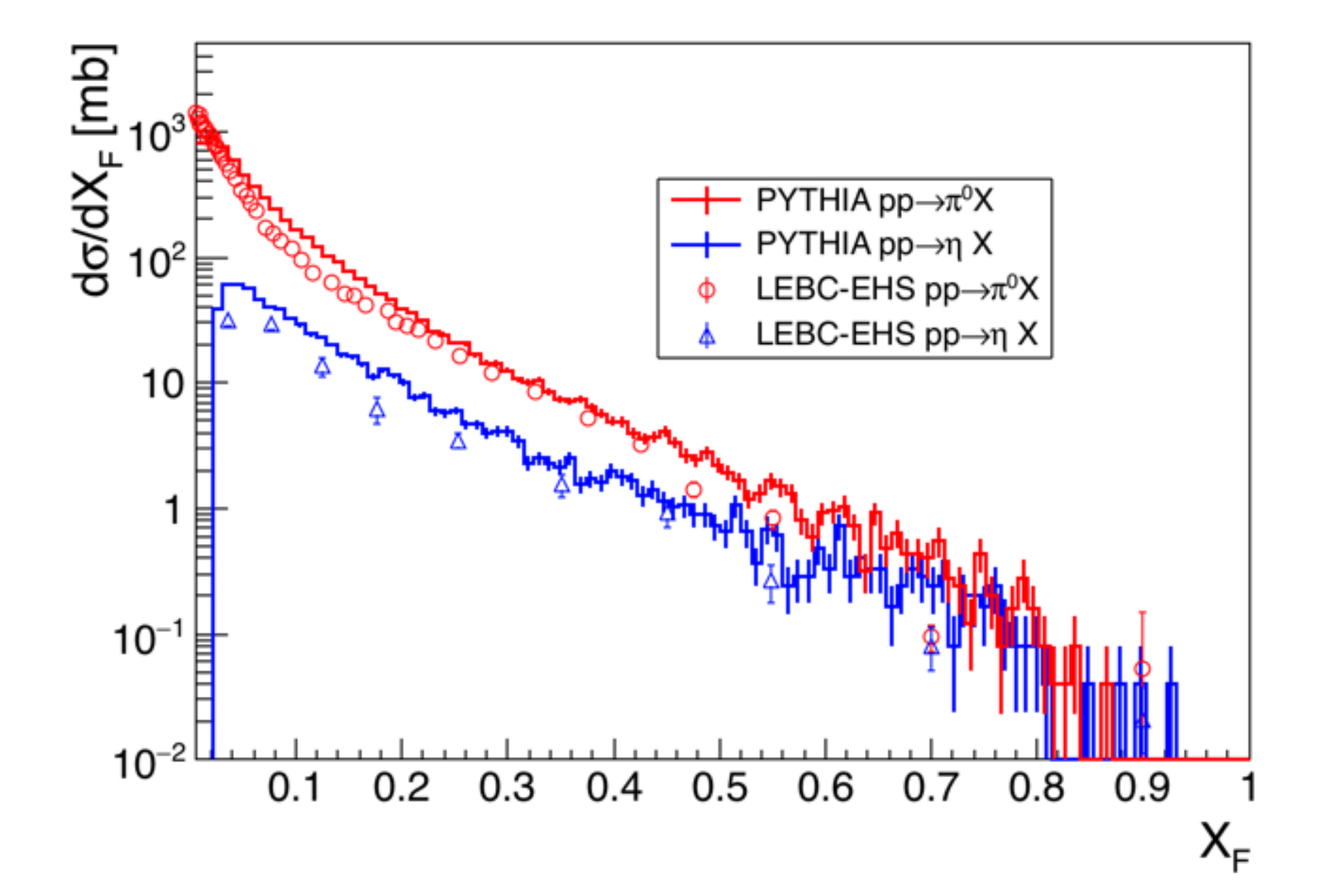}
  \hspace*{1cm}
 \caption{\label{fig:trigge}
Left: Number of charged particles in the acceptance of the trigger chambers in the LEBC-EHS experiment as a function of the Feynman $X_F$ of a co-emitted $\pi^0$ meson, as evaluated using a PYTHIA simulation: the red line indicates the average values for each $X_F$ bin. Right: Data-MC comparison of the differential cross-section of neutral pions and $\eta$ mesons produced from 400~GeV protons, after accounting for the trigger efficiency in the LEBC-EHS setup.}
\end{center}
\end{figure}

  A careful review of~\cite{AguilarBenitez:1991yy} shows that
 a comprehensive data-MC comparison for neutral
 particles demands modelling and accounting for the trigger condition
 in the Monte Carlo output. In the LEBC-EHS data, the trigger requirement was to have 3
 or more hits from charged particles in dedicated forward tracking stations 
 right downstream of the hydrogen bubble-chamber target. 
 As stated in~\cite{AguilarBenitez:1991yy}, the trigger efficiency is as low as 2\% when only two charged particles are emitted in acceptance and 
 trigger inefficiencies may alter significantly the differential cross section at large $X_F$. We modelled this effect while analyzing the PYTHIA output~\footnote{The expected distribution of the interaction point has been considered as well.} and found that the number of charged particles emitted in the acceptance of the trigger chambers strongly depends on the $X_F$ of a co-emitted $\pi^0$, as shown in the left panel of Fig.~\ref{fig:trigge}. 
 After correcting for the trigger efficiency, the differential cross section significantly changes, as shown in the right-hand side of 
 Fig.~\ref{fig:trigge}.
 The data-MC ratio within large-$X_F$ bins of sufficient statistics saturates
 at $\sim 0.8$, and one can safely use the entire $X_F$ range for sensitivity studies.
 The data-MC ratio for charged pions and $\eta$ mesons as a function of $X_F$ is higher than 0.8 for any value of $X_F$. 

 Checking the results of Fig.~10 in~\cite{Dobrich:2019dxc} for
 a number of of benchmark points we verify that the sensitivity
 of proposed and past searches does not change very visibly
 in the double-log presentation
 such that the conclusions of~\cite{Dobrich:2019dxc} 
 remain largely unchanged.

\end{document}